\documentstyle[floats,multicol,aps,epsf,prb]{revtex}

\newcommand\relbd{\mathrel{{\bf\smash{{\phantom- \above1pt \phantom-
}}}}}
\newcommand\ltdash{\raise-1.8pt\hbox{$\scriptscriptstyle |$}}
\newcommand \beq  {\begin{equation}}
\newcommand \eeq  {\end{equation}}
\newcommand \bea {\begin{eqnarray} }
\newcommand \eea {\end{eqnarray}}
\newcommand\joinreldex{\mathrel{\mkern-9mu}}
\newcommand\joinrelwx{\mathrel{\mkern-6mu}}\def\up{\uparrow}
\newcommand\gtappr{{{\lower4pt\hbox{$>$} } \atop \widetilde{ \ \ \ }}}
\newcommand\ltappr{{{\lower4pt\hbox{$<$} } \atop \widetilde{ \ \ \ }}}
\newcommand\si{{\sigma}}
\newcommand\rarrow{{\rightarrow}}
\newcommand\dg{{^{\dag}}}

\def\3he{{$^3${\rm He}}}

\def\slD{\raise.15ex\hbox{$/$}\kern-.57em\hbox{$D$}}
\def\dsl{\raise.15ex\hbox{$/$}\kern-.57em\hbox{$\Delta$}}
\def\slp{{\raise.15ex\hbox{$/$}\kern-.57em\hbox{$\partial$}}}
\def\nsl{\raise.15ex\hbox{$/$}\kern-.57em\hbox{$\nabla$}}
\def\sla{\raise.15ex\hbox{$/$}\kern-.57em\hbox{$\rightarrow$}}
\def\slla{\raise.15ex\hbox{$/$}\kern-.57em\hbox{$\lambda$}}
\def\gtwid{\raise.3ex\hbox{$>$\kern-.75em\lower1ex\hbox{$\sim$}}}
\def\ltwid{\raise.3ex\hbox{$<$\kern-.75em\lower1ex\hbox{$\sim$}}}

\def\12{{1\over2}}

\def\part{\partial}

\def\bk{{\bf k}}

\def\bethlogo{\vbox{\bf \line{\hrulefill} 
    \kern-.5\baselineskip 
    \line{\hrulefill\phantom{ ELIZABETH A. MASON }\hrulefill} 
    \kern-.5\baselineskip 
    \line{\hrulefill\hbox{ ELIZABETH A. MASON }\hrulefill} 
    \kern-.5\baselineskip 
    \line{\hrulefill\phantom{ 1411 Chino Street }\hrulefill} 
    \kern-.5\baselineskip 
    \line{\hrulefill\hbox{ 1411 Chino Street }\hrulefill} 
    \kern-.5\baselineskip 
    \line{\hrulefill\phantom{ Santa Barbara, CA 93101 }\hrulefill} 
    \kern-.5\baselineskip 
    \line{\hrulefill\hbox{ Santa Barbara, CA 93101 }\hrulefill}
    \kern-.5\baselineskip 
    \line{\hrulefill\phantom{ (805) 962-2739 }\hrulefill} 
    \kern-.5\baselineskip 
    \line{\hrulefill\hbox{ (805) 962-2739 }\hrulefill}}}
\def\lisalogo{\vbox{\bf \line{\hrulefill} 
    \kern-.5\baselineskip 
    \line{\hrulefill\phantom{ LISA R. GOODFRIEND }\hrulefill} 
    \kern-.5\baselineskip 
    \line{\hrulefill\hbox{ LISA R. GOODFRIEND }\hrulefill} 
    \kern-.5\baselineskip 
    \line{\hrulefill\phantom{ 6646 Pasado }\hrulefill} 
    \kern-.5\baselineskip 
    \line{\hrulefill\hbox{ 6646 Pasado }\hrulefill} 
    \kern-.5\baselineskip 
    \line{\hrulefill\phantom{ Santa Barbara, CA 93108 }\hrulefill} 
    \kern-.5\baselineskip 
    \line{\hrulefill\hbox{ Santa Barbara, CA 93108 }\hrulefill}
    \kern-.5\baselineskip 
    \line{\hrulefill\phantom{ (805) 962-2739 }\hrulefill} 
    \kern-.5\baselineskip 
    \line{\hrulefill\hbox{ (805) 962-2739 }\hrulefill}}}

\def\low#1{\lower.5ex\hbox{${}_#1$}}
\def\ltwid{\raise.3ex\hbox{$<$\kern-.75em\lower1ex\hbox{$\sim$}}}

\def\psl{\raise.15ex\hbox{$/$}\kern-.57em\hbox{$\partial$}}
\def\partt{\raise.15ex\hbox{$\widetilde$}{\kern-.37em\hbox{$\partial$}}}
\def\parts{\raise.15ex\hbox{$/$}{\kern-.6em\hbox{$\partial$}}}
\def\nablas{\raise.15ex\hbox{$/$}{\kern-.6em\hbox{$\nabla$}}}
\def\oprod{\hbox{$\rm O$}{\kern -0.8em\hbox{$\Pi$}}}
\def\partw#1{\raise.15ex\hbox{$/$}{\kern-.6em\hbox{${#1}$}}}

\def\si{{\sigma}}

\def\gtappr{{{\lower4pt\hbox{$>$} } \atop \widetilde{ \ \ \ }}}
\def\ltappr{{{\lower4pt\hbox{$<$} } \atop \widetilde{ \ \ \ }}}

\def\topppageno1{\global\footline={\hfil}\global\headline
={\ifnum\pageno<\firstpageno{\hfil}\else{\hss\twelverm --\ \folio
\ --\hss}\fi}}

\def\toppageno2{\global\footline={\hfil}\global\headline
={\ifnum\pageno<\firstpageno{\hfil}\else{\rightline{\hfill\hfill
\twelverm \ \folio
\ \hss}}\fi}}

\def\relbd{\mathrel{{\bf\smash{{\phantom- \above1pt \phantom-
}}}}}

\def\ltdash{\raise-1.8pt\hbox{$\scriptscriptstyle |$}}
\def \ra{\rangle}

\def\s{\sigma}

\def\dg{{^
{\dag}}}

\def\ra{\rangle}

\def\1{{\bf 1}}
\def\2{{\bf 2}}

\def\Sfj{{\bf S}_{j}}
\def\rarrow{\rightarrow}

\def\Tr{{\rm Tr}}
\def\dt{\partial _\tau}

\def\vk{\vec k}

\def\ell{{\it l } {\rm n}}

\def\si{\sigma}

\def\cx2{\sqrt{c^2_x+c^2_y}}

\def\gkk{\gamma _{\vec k}}
\def\gk2{\gkk ^2}
\def\dw{\downarrow}
\def\up{\uparrow}
\def\gtappr{{{\lower4pt\hbox{$>$} } \atop \widetilde{ \ \ \ }}}
\def\ltappr{{{\lower4pt\hbox{$<$} } \atop \widetilde{ \ \ \ }}}

\def\pbar{{\partial\kern-1.2ex\raise0.25ex\hbox{/}}}

\def\up{\uparrow}
\def\dw{\downarrow}

\def\s{\sigma}

\def\dg{{^{\dag}}}

\def\ra{\rangle}

\def\1{{\bf 1}}
\def\2{{\bf 2}}

\def\rarrow{\rightarrow}

\def\Tr{{\rm Tr}}
\def\dt{\partial _\tau}

\def\vk{\vec k}

\def\ell{{\it l } {\rm n}}

\def\si{\sigma}

\def\cx2{\sqrt{c^2_x+c^2_y}}

\def\gkk{\gamma _{\vec k}}
\def\gk2{\gkk ^2}
\def\gtappr{{{\lower4pt\hbox{$>$} } \atop \widetilde{ \ \ \ }}}
\def\ltappr{{{\lower4pt\hbox{$<$} } \atop \widetilde{ \ \ \ }}}

\def\thickra{\hbox{\raise0.2pt\hbox{{$\bf >\mkern-13mu>\mkern-13mu>$}}}}
\def\thickrarrow{\hbox{\raise0.28pt\hbox{{$\bf >\mkern-13mu>\mkern-13mu>$}}}}

\begin{document}
\draft
\twocolumn[\hsize\textwidth\columnwidth\hsize\csname @twocolumnfalse\endcsname
\title{Co-operative Kondo Effect in the two-channel Kondo Lattice}
\author{  P. Coleman$^1$ , A. M. Tsvelik$^2$, N.
Andrei$^1$ and H. Y. Kee$^3$}
\address{$^1$
Serin Laboratory, Rutgers University,
P.O. Box 849, Piscataway, NJ 08855-0849, USA.}
\address{$^2$ Department of Physics,
Oxford University
1 Keble Road
Oxford OX2 3NP, UK
}
\address{$^{3}$ Department of Physics, University of California at Los Angeles,
Los Angeles, California 90024}
\maketitle
\begin{abstract}

We examine how the properties of a single-channel Kondo lattice model
are modified by additional screening channels.  Contrary to current
wisdom, additional screening channels appear to constitute a relevant
perturbation which destabilizes the Fermi liquid.  This instability
occurs involves two stages.  When a heavy Fermi surface develops, it
generates zero modes for Kondo singlets to fluctuate between screening
channels of different symmetry, producing a divergent composite pair
susceptibility.  Additional screening channels couple to these
divergent fluctuations, promoting an instability into a superconducting 
state with long-range composite order. We discuss possible implications
for heavy fermion superconductivity. 

\end{abstract}
\vskip 0.2 truein
\newpage
\vskip2pc]

\section{Introduction}

One of the remarkable properties of localized  magnetic moments is their
ability to transform the electronic properties of their host. 
These effects are dramatic in heavy fermion compounds.
\cite{heavy,steglich} 
Since the mid-seventies, several hundred heavy fermion compounds have been
discovered,  characterized by a dense lattice of
magnetic rare-earth, or actinide ions immersed in
a conducting host.  These materials
by-pass the normal development of ordered
antiferromagnetism to form a new kind of electron fluid. 
The resulting metallic state contains quasiparticles 
with effective masses up to a thousand times greater than a bare
electron. For example, in  $CeCu_6$\cite{cecu6} the presence of only 
14\% Cerium in the copper host increases the effective mass of
the electrons by a factor of 1600.   

In a small handful of heavy fermion compounds,
the heavy electron fluid becomes
superconducting.\cite{heavy} Local moments, 
usually harmful to superconductivity
actually participate in this superconducting
condensation process and   a
significant fraction of the local moment entropy
is quenched as part of the condensation process. 
In $UBe_{13}$ for example,  the spin-condensation
entropy is about
$0.2k_B ln2$ per spin.\cite{ube13ref}
One of the great challenges is to understand
how microscopic
order parameter in these systems 
involves the spin operators of the local moments.

The
concept of ``composite pairing'', where a Cooper
pair and local moment form a bound-state
combination that collectively condenses may
provide a way to address this problem. 
\cite{emery,balatsky,miranda,bonca,poilblanc,zachar}
A composite ``triplet'' involves a
bound-state between  a spin and singlet Cooper pair,
\bea
\pmb{$\Lambda$}_t(x)= \langle \Psi_{N-2}\vert
\psi_{\downarrow}(x)\psi_{\uparrow}(x)
{\bf S}(x)\vert \Psi_{N}\rangle,
\eea 
but a composite singlet involves
a triplet and a spin-flip,
\bea
\Lambda_s (x)= \langle \Psi_{N-2}\vert
\psi_{1\downarrow}(x)
\psi_{2\downarrow}(x){\bf S}^+(x)\vert  \Psi_{N} \rangle.
\eea
where 
``1'' and ``2'' refer to two conduction electron channels. Such composite
order parameters were originally considered in the context of odd-frequency
pairing\cite{emery,balatsky,miranda} but more recent work has emphasized
that composite order may co-exist with BCS
pairing in cases where the spin plays a central role
in the condensation process.\cite{bonca,poilblanc}
Unfortunately, we know very little about how such composite
pairing might come to pass. 
A divergent composite singlet susceptibility 
is known to occur in the symmetric two-channel
Kondo impurity model,\cite{emery} and 
more recent studies suggest that a large composite susceptibility may persist into the 
two-channel Kondo lattice. \cite{jarrell}

In this paper we introduce a model for  heavy fermion behavior 
where the local moments couple to a
{\sl single} conduction band via  two orthogonal 
scattering  channels.
We find that when two 
scattering channels of the same parity
share a common  Fermi surface, 
{\sl constructive} interference
develops between the 
channels. 
The scattering of electrons in the Kondo effect is described
by an SU(2) matrix ${\cal V}_{\Gamma}$, ($\Gamma = 1,2$) 
associated with each channel. A key result of our paper relates
the composite order to the gauge invariant interference term
between these two matrices:
\bea
{\cal V}\dg_2{\cal V}_1 = - \frac{J_1J_2}{2}
\left[
\matrix{
F &  \Lambda\cr
- \Lambda \dg & F\dg 
}
\right],
\eea
where
\bea
F &= &\psi\dg_1 \pmb{$\sigma$} \psi_2 \cdot {\bf
S},\cr 
{\Lambda} &= &\psi_1 ( i \sigma_y)\pmb{$\sigma$}
\psi_2
\cdot {\bf S},\eea
represent the singlet composite order in the
particle-hole, and particle-particle channels
respectively, and $J_1$ and $J_2$ describe the
Kondo coupling constants in the two channels. 
In a single impurity, 
the
Kondo effect in the stronger channel,
suppresses any Kondo effect
in weaker channels.
\cite{blandin,coxrev} 
A key feature of our lattice mechanism,
is that channel interference co-operatively enhances the
Kondo effect in the weaker channel, driving the development
of composite pairing for arbitrarily weak second-channel
coupling. 
\begin{figure}[tb]
\epsfysize=2.6in
\centerline{\epsfbox{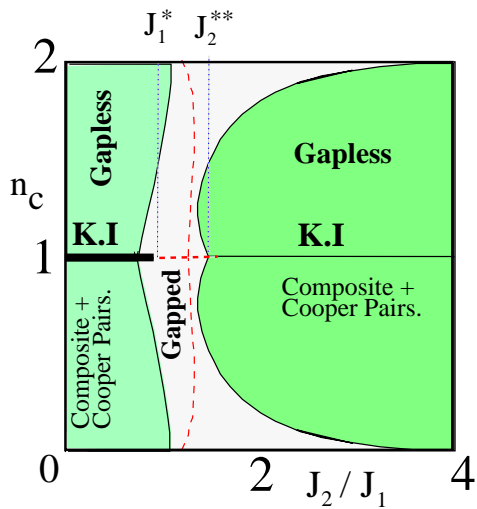}}
\vskip 0.3truein
\protect\caption{Mean-field phase diagram for a two-dimensional
two-channel Kondo lattice with an s-wave and d-wave interaction
channel. $n_c$ denotes the filling of the conduction band.
$J_2/J_1$ is the ratio of coupling constants in the two channels.
The phase diagram was computed for a tight-binding model,
keeping $max(J_1,J_2) = 4t$, the band-width.
``K.I ''  denotes  ``Kondo insulator'' phases, that
exist at half-filling, but which undergo a superconductor-insulator
transition at a critical value of $J_2/J_1$.
The lightly shaded region is dominated
by composite pairing, and there is a gap for quasiparticle excitations. 
Along the dotted lines, the conduction electrons are entirely unpaired,
and a pure composite pair condensate is formed.  In the darkly shaded
regime, Cooper and composite pairs co-exist, and a gapless anisotropic
superconductor is formed. 
}
\label{Fig5q}
\end{figure}

The
development of phase coherence between the two
channels  is signalled by the condensation  of
composite singlet pairs  at a new
temperature scale 
\bea
T_c \sim \sqrt {T_{K1}T_{K2}},
\eea
where $T_{K1}$ and $T_{K2}$ are the Kondo temperatures of the
primary and secondary channels respectively. 
The underlying gap symmetry of the quasiparticles in this
new superconducting phase reflects the
interference phenomena and is 
is given by a product of the form factors $ \Phi_{1 {\bf k}}$
and $\Phi_{2 {\bf k}}$ from each channel:
\bea
\Delta_{\bf k}  = \Delta_o \Phi_{1 {\bf k}}\Phi_{2 {\bf k}}.
\eea

In the typical composite paired state, composite pairs 
co-exist with Cooper pairs, as envisaged in the works
of  Bonca and Balatsky and also Poilblanc.\cite{bonca,poilblanc}
One of the novel  features of this mechanism,
is that it permits both gapless, \underline{and}
gapfull anisotropic superconductivity. 
In the region where the coupling constants
$J_1$ and $J_2$, for the two channels,  are of comparable
strength, the nodes in the excitation spectrum
gravitate to the center of the unit cell, where they mutually
annihilate to produce a gapped phase (Fig. \ref{Fig5q}., Fig. \ref{Fig7}.).  
In this phase, the BCS order parameter is very small, and actually
vanishes along a line in the 
phase diagram.

At half filling, a Kondo lattice generally forms a Kondo insulating
phase.\cite{tsunetsugu} With the addition of a second-channel coupling, 
we find that at a critical
ratio of coupling constants, there is a second-order transition
from the Kondo insulating phase into a pure
condensate of superconducting pairs. 
This leads to a phase diagram, where a first order line
representing the Kondo insulator terminates at a superconducting-insulating
transition, as illustrated in Fig. (\ref{Fig5q}).

An brief description of our work in this area has already been published. 
\cite{early}
This paper is intended to provide a 
detailed account and discussion of the co-operative
two-channel Kondo effect. 

\section{Multi-channel Scattering Effects in interacting Kondo lattices}

The classical approach to heavy fermion physics
involves  local moments which
couple exclusively 
to conduction electron states with the same local f-symmetry.
This assumption derives from the observation
that spin-exchange between the conduction
electrons and the local moments occurs
predominantly via  hybridization in the
f-channel.

However, more careful considerations suggest 
\cite{melnikov,fulde,tsvelik1} 
that electron-electron interactions
can cause new
spin-exchange channels to open up between a local moment and the
conduction sea. There are several mechanisms by
which these new spin exchange channels can open
up, including
\begin{itemize}

\item Vicinity to a quantum critical
point.\cite{melnikov}

\item Interactions in the conduction
sea.\cite{fulde,tsvelik1}

\item Intra-atomic Hunds interactions.\cite{coxphysica}

\end{itemize}
The first mechanism,  identified
long ago in a little-known paper  
by Larkin and Melnikov\cite{melnikov}
may be particularly important for heavy fermion 
systems which lie at the brink of magnetism.  
Larkin and Melnikov studied the single
impurity Kondo effect 
in the vicinity of a magnetic 
quantum phase transition, where the local 
local moment polarizes electrons at increasingly
greater distances. The critical magnetic
order thereby induces  the spin
to scatter electrons in a large number of
 angular momentum channels up to a maximum
value $l_o\sim
k_F
\xi$, where $\xi$ is the spin correlation length.
The large screening cloud causes 
the matrix element for spin exchange to become
\bea
J\rightarrow 
J_{{\bf k},{\bf k'}} = J \chi({\bf k}- {\bf k'}),
\eea
where $\chi$ is the 
 strongly momentum dependent susceptibility of
the magnetic host.
When decomposed into partial wave states, they
found
that this led to a Kondo coupling to electrons in all channels
with angular momentum 
$
l\le l_o = k_F \xi
$.

\begin{figure}[tb]
\epsfxsize=3.0in 
\centerline{\epsfbox{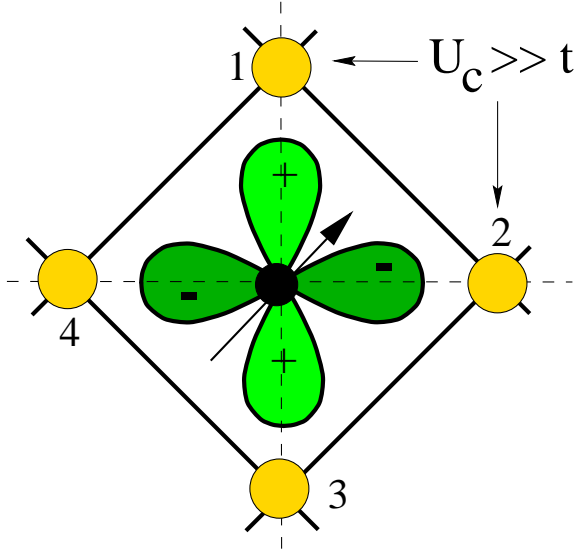}}
\vskip 0.4truein
\protect\caption{Magnetic moment in an interacting environment.
Localized electron at center of plaquet hybridizes in the $d_{xy}$-channel
with  nearby atoms.  The on-site interaction at each atomic
site $U_c$ is taken to be far larger than the electron band-width $t$.
}
\label{fig1}
\end{figure}

More recent work has made it clear that
new spin exchange channels open up whenever 
charge fluctuations are suppressed by interactions
in the conduction sea.\cite{fulde,tsvelik1} 
Consider the situation shown in
Fig \ref{fig1}., where
a local moment hybridizes with nearby orbitals in a 
a d-channel.  The spin-exchange between the local
moment is written
\bea
H_I = J({\bf S} \cdot \Psi\dg_{d\alpha}
\pmb{$\si$}_{\alpha \beta}\Psi_{d\beta}),
\eea
where 
\bea
\Psi^{\dag}_{d\si} =
\frac{1}{2}(c^{\dag}_{1\si}-c^{\dag}_{2\si} +
c^{\dag}_{3\si}- c^{\dag}_{4\si})
\eea
creates an electron in the d-channel. Notice that
the spin exchange  involves processes where the
electrons ``hop and flip'' between
neighboring orbitals.
If large repulsive interactions are present in
the conduction sea, then  an electron can no
longer ``hop and flip'' onto a site that is
already occupied. This restriction
means that creation
operators must be replaced by Hubbard operators,
\bea
c_{j\sigma} \longrightarrow c_{j \sigma}(1 - n_{j -\sigma}) = {\rm X}_{j \sigma}, 
\eea
To see how this
modifies the spin-exchange processes one can use a
Gutzwiller approximation, 
\bea
{\rm X}{^{\dag}}_{j } \pmb{$\sigma$}
{\rm X}_{l}\longrightarrow 
c{^{\dag}}_{j} \pmb{$\sigma$}
c_{l}
\times \left\{
\begin{array}{cr}
1,&(j=l)\cr
(1-x)
,&(j \ne l)\end{array}\right.
\eea
where $x$ is the concentration of carriers per site. This
approximation correctly describes the complete suppression of
hop and flip processes in the 
limit where $x=1$.
With this replacement the transformed Kondo interaction develops 
three new scattering channels, 
\bea
H_I &=&  J_1{\bf S}
\cdot (\Psi\dg_{d} \pmb{$\si$}\Psi_{d})\cr
&+& J_2{\bf S}
\cdot ( \Psi\dg_{s} \pmb{$\si$}\Psi_{s}
+ \Psi\dg_{p_x} \pmb{$\si$}\Psi_{p_x}
+\Psi\dg_{p_y} \pmb{$\si$}\Psi_{p_y}),
\eea
where $J_1= (1 - \frac{3x}{4}) J$, 
$J_2= \frac{x}{4} J$, 
\bea
\Psi^{\dag}_{s\si}\ =
\frac{1}{2}(c^{\dag}_{1\si}+c^{\dag}_{2\si} +
c^{\dag}_{3\si}+ c^{\dag}_{4\si}),
\qquad\hbox{\underline{s-channel}}\cr
\cr
\left.\begin{array}{rcl}
\Psi^{\dag}_{p_y\si}&=&
\frac{1}{\sqrt
{2}}(c^{\dag}_{1\si}-c^{\dag}_{3\si}),\cr
\Psi^{\dag}_{p_x\si}&=&
\frac{1}{\sqrt
{2}}(c^{\dag}_{2\si}-c^{\dag}_{4\si}),
\end{array}
\right\}
\quad\qquad\qquad\hbox{\underline{p-channel}}
\eea
create electrons in the secondary channels.
Electrons in the secondary channels are able
to exchange spin with the local moment even
though they do not hybridize with it. 

In more complex Uranium heavy fermion
systems, intra-atomic interactions 
play a vital role in opening up second-channel
couplings. \cite{coxphysica} In Uranium atoms,
the Hunds interactions have the effect of
suppressing  fluctuations in  the ``shape'' of the
localized orbital, so that electrons scattering
off a localized orbital tend to exchange spin,
whilst preserving their orbital quantum numbers.
In a tetragonal crystal for example where the
low lying state of the $f^2$ ion is a magnetic
non-Kramers  doublet\cite{coxphysica,koga}
\bea
\vert \pm\rangle = \alpha \vert \pm 1 \rangle
+\beta \vert \mp 3 \rangle,
\eea
spin fluctuations within this doublet involve
the exchange of spin with  conduction electrons
in two different ``shape'' channels,
with equal Kondo coupling constants.

\section{Two-channel Kondo Lattice Model}

This discussion motivates us to examine how additional spin exchange
channels might modify the physics of a Kondo lattice. 
To this end, we shall consider a
Kondo lattice model  where two orthogonal scattering channels 
dominate the spin exchange process:
\bea 
H &=& \sum \epsilon_{{\bf k}\sigma}
c{^{\dag}}_{{\bf k}\sigma} c_{{\bf k}\sigma}
+ \sum_{\Gamma j}  J_{\Gamma} \psi{^{\dag}}_{\Gamma j}
\pmb{$\sigma$} \psi_{\Gamma j} \cdot {\bf S}_j,
\eea
where 
$
\psi^{\dag}_{\Gamma j} = (
\psi^{\dag}_{\Gamma j\uparrow},\ 
\psi^{\dag}_{\Gamma j\downarrow})
$, ($\Gamma = 1, \ 2$)
is a two component spinor
\bea
\psi{^{\dag}}_{\Gamma j\sigma } =N_s^{-\frac{1}{2}}
\sum_{\bf k} \Phi_{\Gamma {\bf k}}
c{^{\dag}}_{{\bf k}\sigma} e^{-i {\bf k} \cdot
{\bf R}_j}, 
\eea
that creates an electron at site j in one of two orthogonal
Wannier states,  with
form-factor
$\Phi_{\Gamma{\bf k}}$. Here $N_s$ is the number of sites.
We shall show that channel interference becomes strong 
when two channels have the same spatial parity. 

\begin{figure}[tb]
\epsfxsize=3.5 in 
\centerline{\epsfbox{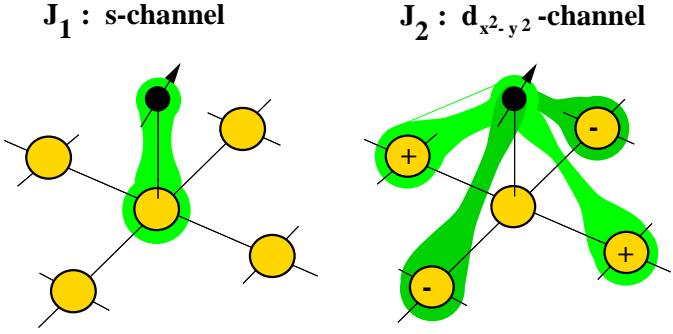}}
\vskip 0.1truein
\protect\caption{Illustrating spin coupled to electrons via
an s, and a $d_{x^2-y^2}$  channel.
}
\label{fig2a}
\end{figure}
A simple example of our model is a two-dimensional
tight-binding lattice of conduction electrons, 
where
\bea
\epsilon_{\bf k} &=& - 2t (\cos k_x + \cos k_y)- \mu,
\eea
and $\mu$ is the chemical potential, 
interacting with a local
moment at each site in an s and a d-channel,
so that
\bea
\begin{array}{rcll}
\Phi_{1\bf k}&=&1,\qquad &\hbox{(s-channel)}\cr
\Phi_{2\bf k}&=&(\cos k_x - \cos k_y), \qquad &\hbox{($d_{x^2 - y^2}$-channel)}.
\end{array}
\eea
as shown in Fig. \ref{fig2a}.

\begin{figure}[tb]
\epsfxsize=3.5 in 
\centerline{\epsfbox{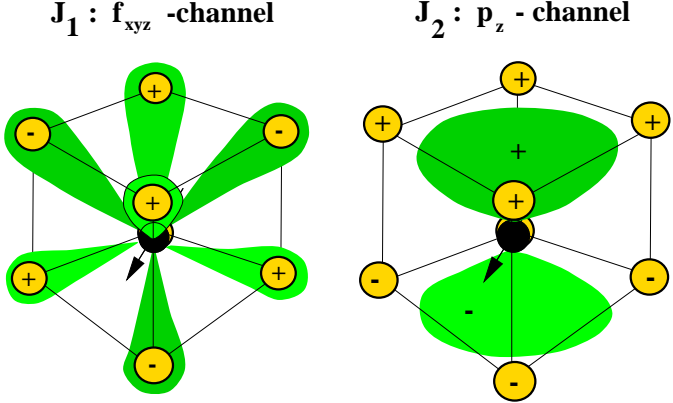}}
\vskip 0.1truein
\protect\caption{Illustrating spin coupled to electrons via
a primary  $f_{xyz}$ and a secondary $p_{z}$  channel.
}
\label{fig2b}
\end{figure}

A slightly more appropriate example would be a three-dimensional
lattice, where
\bea
\epsilon_{\bf k} &=& - 2t (\cos k_x + \cos k_y+ \cos k_z)- \mu,
\eea
with a local moment at the center of each cube of
atoms interacting in a primary $f_{xyz}$ channel, and a secondary
$p_{z}$ channel: 
\bea
\begin{array}{rcll}
\Phi_{1\bf k}&=&\sqrt{8}\sin k_x \sin k_y \sin k_z ,\qquad &\hbox{($f_{xyz}$ -channel)}\cr
\Phi_{2\bf k}&=&\sqrt{2}\sin k_z, \qquad&\hbox{($p_{z}$-channel)}.
\end{array}
\eea
as shown in Fig. 
\ref{fig2b}. 

Unlike 
earlier treatments of two-channel
Kondo problems, our model involves a {\sl single} 
conduction electron band, and there is no
globally conserved ``channel quantum number''.  
In a heavy fermion system, the orbital channels
are locally well defined, but an electron
scattering in one channel at one site, can then
scatter in a different channel at a second site. 
This is important, for
it can lead to interference effects between the
Kondo effect in different channels which are
completely absent in models with an artificial
channel quantum number conservation.  To
illustrate this important point, we shall
contrast the properties of our model with the
channel symmetric ``control model'' 
\bea
H^{C} =
\sum_{\bf k \Gamma \sigma} 
\epsilon_{\bf k}
c{^{\dag}}_{\Gamma{\bf k}\sigma} 
c_{\Gamma{\bf k}\sigma} +
\sum_{\Gamma j} J_{\Gamma}
c{^{\dag}}_{\Gamma j}
\pmb{$\sigma$} c_{\Gamma j} \cdot {\bf S_j}.
\eea
where now
\bea
c{^{\dag}}_{\Gamma j\sigma } ={N_s}^{-\frac{1}{2}}
\sum_{\bf k} 
c{^{\dag}}_{\Gamma {\bf k}\sigma} e^{-i {\bf k}
\cdot {\bf R}_j}, 
\eea
$ (\Gamma = 1, \ 2)$.
In the control, electrons in different channels
do not interfere, and we shall show that this
prevents the development of composite pairing.  

\section{Composite pairing instability of the one-channel Kondo lattice}

To examine the effect of second-channel couplings, we introduce the
composite operator \bea \Lambda\dg= \sum_j
i\psi{^{\dag}}_{1j}\pmb{$\sigma$} \sigma_2
\psi{^{\dag}}_{2j}\cdot {\bf S_j}. \eea 
This operator  transfers singlets between
channels by adding a triplet and
flipping the local moment.  To see how this works, consider
a single site, where
\bea
\vert \Psi_{s1}\ra &=& \frac{1}{\sqrt{2}}\biggl[
\psi\dg_{1\up}\vert \dw \rangle -
\psi\dg_{1\dw}\vert \up \rangle
\biggr],\cr
\vert \Psi_{s2}\ra &=& \frac{1}{\sqrt{2}}\biggl[
\psi\dg_{2\up}\vert \dw \rangle -
\psi\dg_{2\dw}\vert \up \rangle
\biggr],
\eea
represent Kondo singlets in channels one, and two respectively.
The action of the single site composite operator 
$\Lambda\dg = i\psi{^{\dag}}_{1}\pmb{$\sigma$} \sigma_2
\psi{^{\dag}}_{2}\cdot {\bf S}$ is 
as follows:
\bea
\Lambda\dg \vert \Psi_{s1}\ra &=& 2 \psi\dg_{1\dw} \psi\dg_{1\up} \vert
\Psi_{s2}\ra,\cr
\Lambda\dg \vert \Psi_{s2}\ra &=& 2 \psi\dg_{2\up} \psi\dg_{2\dw} \vert
\Psi_{s1}\ra,
\eea
showing that the composite operator $\Lambda\dg$ 
transfers a Kondo singlet between channels, 
leaving an electron pair behind in the channel formerly occupied
by a Kondo singlet.

We now show how
channel interference in the one band model causes 
the susceptibility of this composite operator  to
develop a BCS-like divergence in the Fermi liquid
ground-state.
Suppose that $J_2=0$ and $J_1$ is sufficiently large
for a Kondo effect to develop in channel one. 
In the corresponding Fermi liquid ground-state $\vert \Phi\ra$,
the composite pair susceptibility
is given by
\bea
\chi_{\Lambda}
=\sum _{\lambda} \left\{
\biggl(
\frac
{\langle \Phi \vert \Lambda ^{\dag} \vert \lambda\rangle
\langle \lambda \vert \Lambda \vert \Phi \rangle }
{E_{\Phi}- E_{\lambda}}
\biggr)+ ( \Lambda \rightleftharpoons \Lambda ^{\dag})
\right\}.
\eea
To evaluate the matrix elements appearing 
in this expression, we need to decompose
the composite operator in terms of
quasiparticle
operators. The essence of the Kondo effect is
the development of Fermionic bound-states between
the local moments, and the conduction electrons.
At low energies, the operator
$
(
{\bf S_j} \cdot 
\pmb{$\sigma$}_{\alpha \beta} 
)
\psi_{1 \beta}
$
then behaves as a single bound-state fermion, represented by
the  contraction
\bea
(\mathrel{\mathop{
 {\bf S}_j\cdot
{\pmb{$\sigma$}}_{\alpha \beta} 
) 
\psi_{1\beta}(j) }
^{\displaystyle 
 \ltdash
\joinreldex
\relbd
\joinrelwx\relbd
\joinrelwx\relbd\joinrelwx\relbd
\joinrelwx\relbd
\joinrelwx\relbd
\joinreldex
 \ltdash
{\phantom{_{1\beta}(j)}}
}}
=
{\bar z} f_{j\alpha}.
\eea
where  $\bar z$ is the amplitude for bound-state formation.  
By making this contraction, we imply that in all 
matrix elements between low-lying excitations 
$\vert a\rangle$ and  $\vert b \rangle $ of
the  Fermi liquid,  $(
{\bf S_j} \cdot 
\pmb{$\sigma$}_{\alpha\beta} 
)
\psi_{1 \beta}
$ can be replaced by a Fermi operator, as follows:
\bea
\langle a \vert 
(\mathrel{\mathop{
 {\bf S}_j\cdot
{\pmb{$\sigma$}}_{\alpha \beta} 
) 
\psi_{1\beta}(j) }
^{\displaystyle 
 \ltdash
\joinreldex
\relbd
\joinrelwx\relbd
\joinrelwx\relbd\joinrelwx\relbd
\joinrelwx\relbd
\joinrelwx\relbd
\joinreldex
 \ltdash
{\phantom{_{1\beta}(j)}}
}}
\vert b \rangle =
\bar z\langle a\vert 
 f_{j\alpha}\vert b \rangle .
\eea
It is the contraction of the 
exchange term which gives rise to
a resonant hybridization between f and conduction electrons
\bea
J_1[
\psi{^{\dag}}_{1j}
(
\mathrel{
\mathop{
{\bf S_j}
\cdot \pmb{$\sigma$}
)
\psi_{1j}
}^
{\displaystyle 
 \ltdash
\joinreldex
\relbd
\joinrelwx\relbd
\joinrelwx\relbd
\joinreldex
 \ltdash
{\phantom{_{1j}}}
}}
+ {\rm H. c.}]
=
J_1{\bar z} 
[\psi{^{\dag}}_{1j}
f_{j} + {\rm H. c.}].
\eea
so that 
at low energies, the Kondo  Hamiltonian can
be replaced by an effective Anderson model.

The low energy eigenstates of the one channel
Kondo lattice model
are then   
an admixture of electron
and composite fermion
$
a_{{\bf k}\sigma} = \cos \delta_{{\bf k}} c_{{\bf k}\sigma} + 
\sin \delta_{{\bf k}} f_{{\bf k}\sigma},
$
with  Hamiltonian 
$
H^*=\sum_{{\bf k} \sigma} E_{{\bf k}} a {^{\dag}}_{{\bf k}\sigma} a_{{\bf k}\sigma}
$. The volume of the  Fermi surface now
counts both the conduction and  composite
f-electrons.\cite{martin,millis,affleck2} 
In the one band model,  the conduction and
composite f- electron share a single Fermi
surface and  they may be
decomposed as follows
\bea
c_{{\bf k}\sigma} &=& \cos \delta_{{\bf
k_F}} a _{{\bf k}\sigma} + \dots\cr
f_{{\bf k}\sigma} &=& \sin \delta_{{\bf k_F}} a
_{{\bf k}\sigma} + \dots \label{decomp}
\eea
where the high energy
components that do not affect the low-energy
matrix elements have been omitted . Near the Fermi surface the
scattering is resonant and 
$\delta_{{\bf k_F}} \sim \pi/2$. Moreover, the small conduction electron
admixture at the Fermi surface must reflect the symmetry of the 
screening channel, so that 
$\cos \delta_{{\bf k_F}}  \propto \Phi_{1 {\bf
k_F}}$.

We can now apply the
contraction procedure to evaluate the matrix elements of the
composite operator. Let us begin with the
control model.  Applying the contraction procedure
we obtain 
\bea
\langle \lambda \vert\Lambda\dg \vert \Phi\rangle_C& = &
-i\sum_j\langle \lambda \vert 
\mathrel{\mathop{ {\bf S}_j\cdot (
c^{\dag}_{1j}}
^{
\displaystyle
 \ltdash
\joinreldex
\relbd
\joinrelwx\relbd
\joinrelwx\relbd
\joinreldex
 \ltdash{\phantom{^{\dag}_{1j}}}
}}
\pmb{$\sigma$} \sigma_2 c{^{\dag}}_{2j}
)\vert \Phi\rangle \cr
&=&{z}\sum_{{\bf k}, \sigma}\langle \lambda\vert 
\sigma c^{\dag}_{2 {\bf k}\sigma}
f{^{\dag}}_{-{\bf k} -\sigma}
\vert \Phi\rangle .\eea
In the  control model,
$c^{\dag}_{2 \bf k}$  and 
$f^{\dag}_{-\bf k}\sim a^{\dag}_{-\bf k}$
respectively create  light and heavy electrons  on
completely different Fermi surfaces.  The
mismatch between the volume and the
dispersion of the   Fermi surfaces for
channel one and two assures that the excitation
energy
$E_{\lambda} - E_{\Phi} =\epsilon_{\bf k }+
E_{\bf k}$ is always finite:
\bea 
\left.
\begin{array}{rcl}
\langle \lambda \vert \hat \Lambda\dg\vert \Phi
\rangle_C
&\propto&\sum _{{\bf k},
\sigma} \sigma 
\  \langle \lambda \vert c{^{\dag}}_{{\bf
2 k} \sigma} 
 a{^{\dag}}_{-{\bf k} - \sigma}\vert \Phi \rangle
\cr\cr
E_{\lambda} - E_{\Phi} &= &\epsilon_{\bf k}+
E_{\bf k}>0.
\end{array}\right\}\label{pair}
\eea
The channel susceptibility $\chi_{\Lambda}$ is
consequently {\sl finite}. 
We conclude that with perfect channel symmetry,
a small second-channel
coupling is {\sl irrelevant}.

Now let us remove the channel symmetry and return
to the physical model. Now we have
\bea
\langle \lambda \vert\Lambda\dg\vert \Phi\rangle&
= & i\sum_j\langle \lambda \vert 
\mathrel{\mathop{ {\bf S}_j\cdot (
\psi^{\dag}_{1j}}
^{
\displaystyle
 \ltdash
\joinreldex
\relbd
\joinrelwx\relbd
\joinrelwx\relbd
\joinreldex
 \ltdash{\phantom{^{\dag}_{1j}}}
}}
\pmb{$\sigma$} \sigma_2 \psi{^{\dag}}_{2j}
)\vert \Phi\rangle \cr
&=&{z}\sum_{{\bf k}, \sigma}\Phi_{2{\bf k}} \langle \lambda\vert 
\sigma c^{\dag}_{ {\bf k}\sigma} f{^{\dag}}_{-{\bf k} -\sigma}
\vert \Phi\rangle .\eea
Unlike the previous case, this pair creation
operator can be decomposed in terms of
quasiparticles on a single heavy Fermi surface. 
Transforming to
quasiparticle operators using eq. (\ref{decomp}) 
introduces a
factor $\cos ( \delta_{\bf k}) \sin (\delta_{\bf k})
\sim 
\Phi_{1 {\bf k}}$ into the sum, so that
\bea
\left.
\begin{array}{rcl}
\langle \lambda \vert \hat \Lambda\dg\vert \Phi
\rangle
&\propto&\sum _{{\bf k},
\sigma} \sigma 
\  \Phi_{1  -{\bf k}} \Phi_{2
{\bf k}} \langle \lambda \vert a{^{\dag}}_{{\bf
k} \sigma} 
 a{^{\dag}}_{-{\bf k} - \sigma}\vert \Phi \rangle
\cr\cr
E_{\lambda} - E_{\Phi} &= &2 E_{\bf k}.
\end{array}\right\}\label{pair2}
\eea
This relation describes the decomposition of the composite
pair operator in terms of the low-lying quasiparticles.
Notice that the matrix element is proportional to
$ \Phi_{1  -{\bf k}} \Phi_{2
{\bf k}}$, showing that this amplitude
involves an interference between the two
channels. Furthermore, the two
form-factors must must have the same parity, or
the  composite operator vanishes on the Fermi
surface. Since the excitation energy, $2 E_{\bf
k}$ vanishes on the heavy Fermi surface, it
follows that there are now a large number of zero
modes for the transfer of singlets between
channels. 
\begin{figure}[tb]
\epsfxsize=6.0 in 
\centerline{\epsfbox{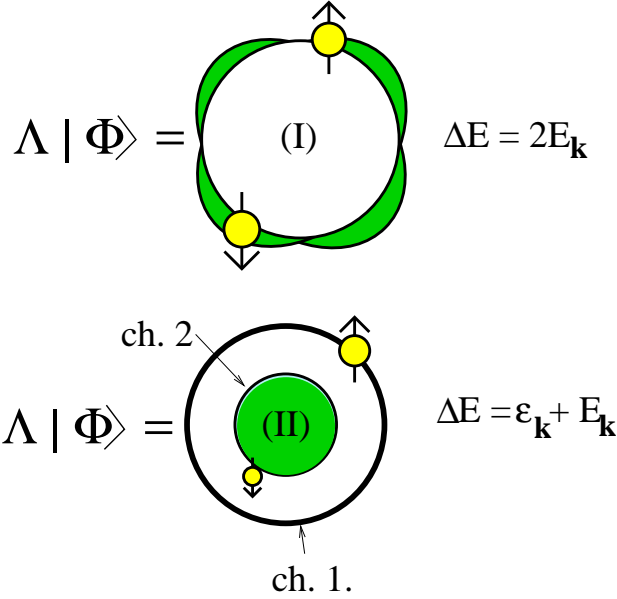}}
\vskip 0.1truein
\protect\caption{Action of composite operator on heavy
Fermi liquid creates: (I) a pair of heavy
fermions (channel interference ) and (II) a heavy and
light electron (channel conservation).
}
\label{fig2}
\end{figure}

It follows that the composite pair susceptibility
$\chi_{\Lambda}$
now contains a singular term.  Substituting
the above results into the general expression
for the composite pair susceptibility, we find
\bea
\chi_{ \Lambda} \propto \sum_{{\bf k}} 
 {(\Phi_{1 {\bf k}}
\Phi_{2{\bf k}})^2\over 2 E_{{\bf k}}} 
\longrightarrow \infty,
\eea
which diverges logarithmically in the
thermodynamic limit. 
We see that once channel symmetry is broken,
the composite pair susceptibility $\chi_{\Lambda}$
is directly proportional to
the BCS pair susceptibility of the
heavy quasiparticles, where the symmetry of the
channel is given by the {\sl product} of the
two screening channels. 
 
\begin{figure}[tb]
\epsfxsize=7.0in 
\centerline{\epsfbox{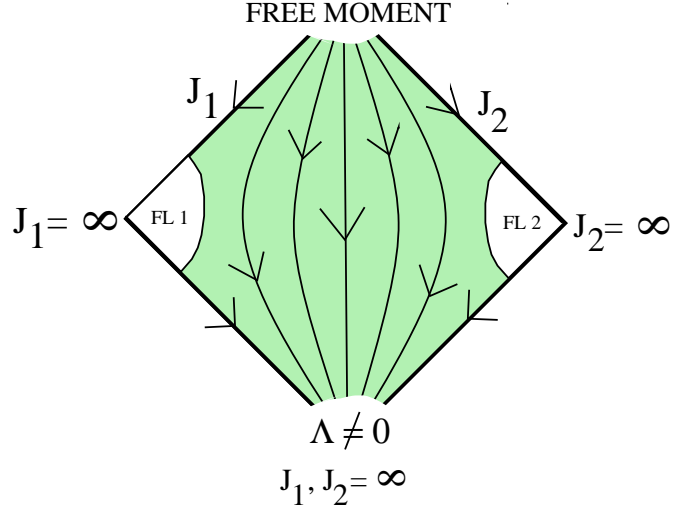}}
\vskip -0.2truein
\protect\caption{
Conjectured renormalization group flows for the co-operative
two-channel Kondo effect. The Fermi liquid formed in channel one or
two is unstable to common a two-channel state with composite order.}
\label{fig3}
\end{figure}

This has immediate consequences for the effect
of a finite $J_2$ on the Fermi liquid ground
state.  Once channel symmetry is
broken, the susceptibility to transfer singlets
by creating composite pairs 
diverges. 
Any finite $J_2$ will
polarize the transfer of singlets into channel
two, thereby coupling
$J_2$ to this divergent susceptibility. 
Thus  the loss
of channel symmetry causes a coupling to a second
channel to become a relevant perturbation.
This will
force $J_2$ to scale to strong-coupling.
A similar
conclusion will hold when $J_2$ is large, and $J_1$ is small.
The simplest way to connect up 
the renormalization flows in the vicinity of the strong-coupling
Fermi liquid fixed points with the flow away from 
the weak coupling fixed point is by hypothesizing the presence of 
a new attractive Kondo lattice fixed point 
that is common to both channels.  (Fig. \ref{fig3})

\section{SU(2) Formalism}

The key to the development of a field theory
for composite pairing, lies in the use of the
Abrikosov pseudo-fermion representation for the
local moments
\bea
\left.
\begin{array}{rcl}
{\bf S}_j   &=& f\dg_{j
\alpha}\left({\pmb{$\si$}\over 2}
\right)_{\alpha
\beta} f_{j \beta},\cr
n_f(j)&=&1.\end{array}\right\}
\eea
Since f-charge fluctuations have been
removed, the KLM is defined within the sub-space 
constrained by the (Gutzwiller) requirement
$n_{f}= 1$ at each site.
The absence of f-charge fluctuations is manifested
as a local SU(2) gauge invariance
of the Heisenberg spin operator
$\Sfj $\cite{affleck} ,
\bea 
f\dg_{j\sigma}\rarrow
\left\{
\begin{array}{c}
e^{i \phi}f\dg_{j \sigma},\cr
\cos \theta f\dg_{j\s} +{\rm sgn} \si sin\theta
f_{j-\s}.  
\end{array}\right.\label{cc}
\eea
To illustrate this feature, consider the spin raising operation
$S_+$.
This process can proceed by first annihilating
a down electron, then creating an up electron, 
written $S^+_{j}=f\dg_{j\up}f_{j\dw}$.
Alternatively, it can proceed by first creating an up electron, forming the
$n_f=2$ state, then annihilating a down electron, written
$S^+_j=-f_{j\dw}f\dg_{j\up}$ . In fact, one can
accomplish the spin raising operation by an
arbitrary linear combination of the above:
\bea 
S^+_j=(\cos \theta f\dg_{j\up} +  sin\theta
f_{j\dw}) (\cos \theta f_{j\dw}-  sin\theta
f\dg_{j\up}).
\eea
In other words, there
is no distinction between a particle
or a hole when all charge fluctuations are removed.
\cite{affleck}
     
The SU(2) symmetry implies that the constraint
$n_f=1$ is actually component of a triplet of local ``Gutzwiller constraints''
\bea
\left.
\begin{array}{c}
f\dg_{j\up}f_{j \up}-
f_{j\dw}f\dg_{j \dw}\cr
f\dg_{j \up}f\dg_{j \dw} \cr
f_{j\dw}f_{j\up}
\end{array}
\right\}
= 0,
\eea
which can be written in the  compact form
\bea
\tilde f\dg_j\pmb{$\tau$}\tilde f_j=0,
\eea
where 
$\tilde f\dg_j=(f\dg_{j \up}, f_{j\dw})$ is a
Nambu spinor, and $\pmb{$\tau$}\equiv(\tau_1,
\tau_2,
\tau_3)$ represents the triplet of Pauli
matrices. 
The first two constraints are particularly
important in any consideration of a paired state, providing 
the main driving force for anisotropic pairing.

The partition function for our model is given by
$
Z= \Tr[ P_G e^{-\beta H}]
$
where $P_G= \prod_j (n^f_{j\uparrow} -n^f_{j\downarrow} )^2$
is the Gutzwiller projection for one f-spin per site. 
Following earlier work, we  rewrite
the Gutzwiller projection as an integral over 
the SU(2) group
\bea 
(n^f_{j\uparrow} -n^f_{j\downarrow} )^2 = \int
d[W_j]\ 
\hat g_j\label{k3},
\eea
where
$
\hat g_j = e^{i f\dg_j W_j f_j}
$  is the SU(2)
operator, 
$W_j= \theta_j \hat {\bf n}_j\cdot
\pmb{$\tau$}$, ( $\theta _j\in
[0 ,2 \pi]$) and
$
d[W] =
{\sin^2 \theta d \theta d \hat n }/(4
\pi ^2)
$
is the Haar measure\cite{zuber} over the
SU(2) group.
Introducing this into the partition function permits 
us to write it as
a path integral
\bea 
Z = \int {\cal D} [f,c,W] e^{-\int_0^{\beta}(
{\cal L}_1+H )d\tau } ,
\eea
where
\bea{\cal L}_1 =
\sum_{\vk} c\dg_{\vk} \dt c_{\vk} +
 \sum_{j } f\dg_j ( \dt - i W_j
) f_j, \label{k5}
\eea
is the Berry phase.

The antiferromagnetic interaction between
the localized moments and the conduction
electrons can be decoupled in the particle-hole
channel, as follows 
\bea
J_{\Gamma}\bigl[\pmb{$\sigma$} _{\Gamma} 
 \cdot
{\bf S}-\frac{1}{2}\bigr] =  -
\frac{J_{\Gamma}}{2}
\{a_{\Gamma} \dg, a_{\Gamma}\},
\eea
where 
\bea
a_{\Gamma}
= \sum_{\si}f\dg_{\si}
\psi_{\Gamma\sigma}.
\eea
The $SU(2)$ gauge symmetry guarantees that
there is in fact, a continuous family of
ways to decouple the interaction. Thus, by
making the transformation $f_{\sigma}\rarrow
\sigma f_{- \sigma}\dg$, we can decouple the
interaction in the Cooper channel, as follows,
\bea
J_{\Gamma}\bigl[\pmb{$\sigma$} _{\Gamma} \cdot
{\bf S}-\frac{1}{2}\bigr] =  -
\frac{J_{\Gamma}}{2}
\{
b_{\Gamma} \dg,  b_{\Gamma}\},
\eea
where
\bea
b_{\Gamma}= \sum_{\si}\sigma f_{-\si}
\psi_{\Gamma\sigma}.
\eea
We now decouple the interaction simultaneously in both channels, by first
writing 
\bea
H_I =  -
\frac{J_{\Gamma}}{4}
\bigl[
\{a_{\Gamma} \dg , a_{\Gamma}\} +
\{b_{\Gamma} \dg,  b_{\Gamma}\} \bigr],
\eea
then decoupling each term 
as follows
\bea
-\frac{J_{\Gamma}}{4}
\{a_{\Gamma} \dg , a_{\Gamma}\}
&\rightarrow&
\bigl[a_{\Gamma}\dg V^{\Gamma} + {\rm h.c }\bigr]
+ \frac{2}{J_{\Gamma}} V^{\Gamma *} V^{\Gamma},\cr
-\frac{J_{\Gamma}}{4}
\{b_{\Gamma} \dg,  b_{\Gamma}\}
&\rightarrow& \bigl[ b_{\Gamma}\dg \Delta^{\Gamma}
+ {\rm h.c }\bigr]  + \frac{2}
{J_{\Gamma}} \Delta^{\Gamma *}
\Delta^{\Gamma}.
\eea

It is convenient at this point, to introduce
a Nambu spinor representation for the
conduction  electrons
\bea 
c_{{\bf k}} =  \left(\matrix{c_{{\bf k} \uparrow} \cr  c\dg _ {-{\bf k}\downarrow} \cr}
\right).\label{f1}
\eea
The corresponding spinor for the localized
electron Wannier states is 
\bea
\tilde \psi_{\Gamma j} =
\left(\matrix{\psi_{\Gamma
\uparrow} \cr  \psi\dg _ {\Gamma\downarrow} \cr}
\right)=
\sum_{\bf k}\Phi_{\Gamma\bf k
 }\left(\matrix{c_{{\bf k} \uparrow} \cr 
{\rm p} c\dg _ {-{\bf k}\downarrow} \cr}
\right)e^{i {\bf k \cdot R}_j},
\eea
where p 
is  the parity of the form-factor
$\Phi_{\Gamma \bf k}= {\rm p}\Phi_{\Gamma -\bf k}
$.
The decoupled interaction can now be
written in the symmetric form (Appendix A ). 
\bea 
J_{\Gamma} 
({\bf S} \cdot \sigma_{\Gamma} )
\rightarrow [\tilde  f\dg
{\cal V}_{\Gamma }\tilde \psi_{\Gamma } + H.c ] +
{1\over  J_{\Gamma} }
 \Tr [{\cal V}\dg_{\Gamma } {\cal V}_{\Gamma} ],
\label{k11}
\eea
where $V_{\Gamma } $ is directly proportional
to an SU(2) matrix $g_{\Gamma }$
\bea
{\cal V}_{\Gamma } = \left[\matrix{
V& \Delta\cr
\Delta^*&-V^{ *}}
\right]^{\Gamma} =i V^{\Gamma }_{o} g_{\Gamma
}  ,\eea
The integration measure for ${\cal V}_{\Gamma}$
is
\bea 
d[{\cal V}_{\Gamma}] = dV_{\Gamma
} dV_{\Gamma
}^* d\Delta_{\Gamma
} d \Delta_{\Gamma
}^*.
\eea

Repeating this decoupling procedure at each
site in the path integral, enables us to  write
\bea 
Z &=& \int {\cal D}[f,c;W,{\cal V}] e^{-
\int_0^{\beta} {({\cal L}_1  +H ) d\tau}},
\cr
H &=&\sum_{\vec k} \epsilon _{\vk}
c\dg_{\vec k}
\tau_3 c_{\vec k}+H_I,
 \cr
H_I&=&\sum_{\Gamma i} \biggl\{[ \tilde f\dg_i
{\cal V}_{\Gamma i }\tilde \Psi_{\Gamma_i} + H.c]
+  { 1\over J_{\Gamma}}
\Tr [ {\cal V} \dg_{\Gamma i} {\cal V}_{\Gamma i}]
\biggr\}. 
\eea
     
\section{Gauge Fixing}

Our model now has the following time dependent
 SU(2) gauge invariance
\bea 
\tilde f_j &\rarrow  & g_j  \tilde f_j , \cr
{\cal V}_{\Gamma j} &\rarrow& g_j {\cal V}_{\Gamma
j},
\cr {W}_j & \rarrow & g_j 
\left( { W}_j  + i\dt \right)g\dg_j,
\label{gtr}
\eea
associated with the absence of f-charge fluctuations. 

When we develop a saddle-point expansion for the
functional integral, we need to deal with the local
zero modes associated with this gauge invariance.
Following standard gauge theory practice, this means
that we need to fix the gauge.  We choose the ``radial
gauge'', where the Kondo matrix ${\cal V}$ is proportional to a unit
matrix in the channel with the largest Kondo coupling constant.
This is the SU(2) analog of the radial gauge used by Read
and Newns in their U(1) treatment of the single channel\cite{read,auerbach} 
Kondo lattice.
Suppose that 
\bea
{\cal V}_1(j) = i v_1({j})h_j,
\eea
where $v_1({j})$ is real and $h_j$ is an SU(2)
matrix.   To fix the gauge, we absorb $h_j$ into
a redefinition of the fields by setting 
$g_j=h\dg_j(\tau)$
and making the gauge
transformation (\ref{gtr}).  In the radial gauge, 
\bea
{\cal V}_{1j}(\tau) &=& i v_{1j}(\tau) \underline{1},
\cr 
{W}_j(\tau) & =& h\dg_j(\tau) 
\left( { W}_j  + i\dt \right)h_j(\tau),
\label{fixed}
\eea
so the formerly static
field $W_j$ is elevated to the status of a 
dynamic field $W_j(\tau)$. 
The measure for the bosonic fields inside the path integral
is now
\bea
d[{\bf W}, {\cal V}]= (v_1)^3 dv_1  d[{\cal V}_2]d^3 {\bf W},
\eea
at each site and time-slice. 

\vskip 0.3 truein
\section{Link Between Composite Order and Channel
Interference}

Under the local SU(2) gauge transformation
$\tilde f_j \rarrow g_j \tilde f_j$, 
(\ref{gtr}) the two 
amplitudes  ${\cal V}_{1j}$ and ${\cal V}_{2j}$
which describe the Kondo effect at site j
transform
in precisely the same way. The only gauge
invariant term we can form from a single
channel is trivially proportional
to the unit matrix:
$
{\cal V}^{\dag} _{\Gamma j}{\cal V} _{\Gamma j}
\propto= \underline{1}
$.
But in a two-channel Kondo problem, 
the interference term 
\bea
 {\cal V}\dg _{2j}{\cal V} _{1j}
\eea
is also gauge invariant, 
since 
$
 {\cal V}\dg _{2j}{\cal V} _{1j} \rarrow {\cal
V}\dg _{1j}g\dg_j g_j {\cal V} _{2j} = {\cal
V}\dg _{2j}{\cal V} _{1j}
$
and for this reason, is expected to have a simple
physical significance. 
To identify the meaning of the interference term
we introduce a source
term into the Hamiltonian that couples to it
\bea
H\rarrow H + 
\sum_j{\rm Tr}
[{\cal V}_{2j}\dg {\cal V}_{1j} \alpha_j + h.c ],
\label{start2}
\eea
where $\alpha_j= \alpha^0_j +
i\pmb{$\alpha $}_j
\cdot {\pmb{$\tau$}} $ is a unitary matrix, with
four real coefficients $(\alpha^0, \pmb{$\alpha
$})$ at each site . If we now
reverse the Hubbard Stratonovich transformation,
by integrating over the fields
${\cal V}_{\Gamma}$ (Appendix B) the Hamiltonian
acquires the additional term 
\bea
H\rarrow H + 
{\rm Tr}
[{\cal M}_j\dg\alpha _j+ h.c ],
\label{start3}
\eea
where now
\bea
{\cal M}\dg_j= 
 - \frac{J_1J_2}{2}
\left[
\matrix{
F &  \Lambda\cr
- \Lambda \dg & F\dg
}
\right]_j,
\eea
and
\bea
F _j&= &\psi\dg_{1j} \pmb{$\sigma$} \psi_{2j}
\cdot {\bf S}_j,\cr 
{\Lambda}_j &= &\psi_{1j}  i
\sigma_y\pmb{$\sigma$}
\psi_{2j}
\cdot {\bf S}_j.\eea
represents the composite order in the
particle-hole, and particle-particle channels
respectively.
By comparing (\ref{start2}) and (\ref{start3}),
we obtain a special relationship between
the inter-channel interference 
and the composite order,
\bea
{\cal V}\dg_{2j} {\cal V}_{1j} 
= 
 - \frac{J_1J_2}{2}
\left[
\matrix{
F &  \Lambda\cr
- \Lambda \dg & F\dg
}
\right]_j.\eea
Notice incidentally that the off-diagonal terms
are odd under interchange of the channel index.

We thus learn that 
{\sl if the Kondo effect develops coherently
in two channels, composite
order develops}.  This enables us to understand
why 
composite order 
develops critical correlations in the
symmetric two-channel Kondo model. \cite{emery}
In a  lattice, true long-range order becomes
possible. 

Let us briefly consider the possible phases
that might develop. If  ${\cal
V}_{\Gamma}$ develops a finite amplitude 
in both channels then the composite order
in the ground-state will have the form
\bea
\left[
\matrix{
\langle \psi \vert F 
\vert \psi \rangle &  \langle \psi
\vert\Lambda\vert \psi \rangle \cr - \langle \psi
\vert\Lambda \dg \vert \psi \rangle &\langle \psi
\vert F\dg\vert \psi \rangle  }
\right]_j=
-\left({2\over J_1J_2}\right)   {\cal V}\dg_{2j}
{\cal V}_{1j} . 
\eea
Suppose the amplitudes of ${\cal V}_{\Gamma}$ are constant, then
in the ``radial gauge''
\bea
{\cal V}_{1j} &=& i v_1\underline{1},\cr
{\cal V}_{2j} &=& i v_2 e^{-i \phi_j {\bf
n}_j\cdot \pmb{$\tau$} },
\eea
where the vector $n_j$ develops a vacuum expectation value.
The composite order matrix is then 
\bea
\left[
\matrix{
\langle \psi \vert F 
\vert \psi \rangle &  \langle \psi
\vert\Lambda\vert \psi \rangle \cr - \langle \psi
\vert\Lambda \dg \vert \psi \rangle &\langle \psi
\vert F\dg\vert \psi \rangle  }
\right]_j=  M_o e^{i \phi_j {\bf
n}_j\cdot \pmb{$\tau$} },
\eea
where $M_o = 2v_1v_2/ (J_1J_2)$.
Two  kinds  of phase are possible:
\begin{itemize}

\item {\bf Composite magnetism} where 
${\bf n }_j= \hat {\bf z}$. In this phase, the
order parameter matrix is diagonal, and
\bea
\langle \psi \vert \psi\dg_{1j} \pmb{$\sigma$}
\psi_{2j}
\cdot {\bf S}_j
\vert \psi \rangle = M_o e^{i \phi_j}.
\eea
This phase breaks time-reversal symmetry,
forming an orbital magnet
where the spin becomes correlated with
electrons in two orbitals.

\item{\bf Composite Singlet pairing} where
$\phi_j = \pi/2$. If  $\hat {\bf n}(x) = \cos
\theta (x)
\hat {\bf x} + \sin \theta (x) \hat {\bf y}$,
whereupon
\bea
\Lambda_s(x_j)
 = i\langle \psi \vert
\psi_{1j}\sigma_y\pmb{$\sigma$}\psi_{2j}\cdot 
{\bf S}_j
\vert \psi \rangle = iM_o e^{-i \theta_j}.
\eea

\end{itemize}   
The second possibility is particularly
interesting, because  the composite pair
susceptibility diverges in the Fermi liquid
phase.  This is the main topic of the 
of the paper.  

\section{Mean Field Theory of the Composite
Paired State}

We now develop a mean-field theory
for the uniform composite paired state.  With this
theory, we show that the two strong-coupling
Fermi liquid phases of our two-channel Kondo
model share a common instability into a phase
with uniform composite order.

We seek a uniform solution, where
all mean-field parameters have no dependence
on position.  In this case 
the mean-field Hamiltonian is most compactly represented in
momentum space, as 
\bea
H_{MF} = \sum_{\bf k}
(\tilde c{^{\dag}}_{\bf k},\ 
\tilde f{^{\dag}}_{\bf k} )
\left[
\matrix{
\epsilon_{\bf k}\tau_3 & {\cal V}{^{\dag}}_{\bf k}
\cr
{\cal V}_{\bf k}& {\bf W} \cdot {\pmb{ $ \tau$}}
}
\right]
\left(
\matrix{
\tilde c_{\bf k}\cr
\tilde f_{\bf k} }\right) 
,\label{matrix}
\eea
where now
\bea
{\cal V}_{\bf k}= {\cal V}^1 {\Phi}_{1{\bf k}}+ {\cal V}^2 
{\Phi}_{2{\bf k}}.
\eea
Strictly speaking, here we should have written the form factors 
as $\Phi_{\Gamma  {\bf k}\tau_3}$,  to take account of the possibility
of an odd-parity scattering channel.  However, provided
both channels have the same parity, we can always
cast ${\cal V}_{\bf k}$ in the above form. For even
parity channels, $\Phi_{\Gamma {\bf k}
\tau_3}= \Phi_{\Gamma {\bf k} }$ directly. For odd-parity
channels, $\Phi_{\Gamma {\bf k}
\tau_3}= \Phi_{\Gamma {\bf k} }\tau_3$, but in this case
the $\tau_3$ can be absorbed by a gauge transformation
$\tilde f_j \rarrow \tau_3 \tilde f_j$, 

To examine uniform pairing, we shall take
\bea
{\cal V}_1&=& i v_1\underline{1},\cr
{\cal V}_2&=&  v_2
{\bf n}\cdot \pmb{$\tau$},\cr
{\cal W}&=& \lambda \tau_3,
\eea
where ${\bf n}= \cos\theta \hat {\bf y}
- \sin \theta \hat {\bf x}$ describes the phase of the 
composite pairing. For convenience we shall take
$\hat{\bf n}=   \hat {\bf y}$, so that
\bea
{\cal V}_{\bf k} = i v_{1 {\bf k} }+ v_{2 {\bf k}}\tau_2,
\eea
where we have introduced the notation
$v_{\Gamma {\bf k } } = v_{\Gamma} \Phi_{\Gamma {\bf k}}$.

Ostensibly, our mean-field theory is that of a 
BCS superconductor, with Hamiltonian described by
\bea
{\cal H}({\bf k})= \left[
\matrix{
\epsilon_{\bf k}\tau_3 & {\cal V}{^{\dag}}_{\bf k}
\cr
{\cal V}_{\bf k}& {\bf W} \cdot {\pmb{ $ \tau$}}
}
\right].
\eea
However, there is one  important distinction:
here the pairing takes place between 
charged conduction electrons
and the neutral f-spins, and is 
merely a manifestation of the
formation of composite pairs.  
For this reason, 
it is actually not possible to say whether the
pairing is channel one, or in channel two. 
In the gauge we have chosen, the 
scattering 
in channel one is  ``normal''  and pairing takes
place in channel two. But suppose 
we make the gauge transformation (\ref{gtr}) with
$g_j = -i \tau_2$, then 
\bea
{\cal V}_{\bf k} = i v_{1 {\bf k} }+ v_{2 {\bf k}}\tau_2
&\longrightarrow& i \tau_2 {\cal V}_{\bf k} = 
  v_{1 {\bf k} }\tau_2-i v_{2 {\bf k}},
\cr&&\cr
{\bf W}= \lambda \tau_3 &\longrightarrow&  i \tau_2{\bf W}
(-i \tau_2) = - \lambda \tau_3,
\eea
which transforms the Hamiltonian to one which is now
pairing in channel {\sl one}, and ``normal'' in channel 
two.  We are forced to recognize 
the superconductivity can not be identified
with either channel, but instead 
derives from a coherence between the two channels. 

Suppose we now integrate out the f-electrons: now we find
that the conduction electron Green function 
has the form 
\bea
{ G}(\kappa)^{-1}
= 
\omega - \epsilon_{\bf k} \tau_3 - \Sigma(\kappa),\label{conduction}
\eea
where $\kappa\equiv({\bf k}, \omega)$ and 
where the self-energy term
\bea
\Sigma(\kappa)= 
{\cal V}\dg_{\bf k}
( \omega -  \lambda  \tau_3 )^{-1}{\cal V}_{\bf k}
\eea
describes the resonant scattering off the quenched
local moments. 
If we expand the self-energy, we see that
it contains both normal and anomalous components
\bea
\Sigma(\kappa)=\Sigma_N(\kappa)+\Sigma_A(\kappa).
\eea
Notice that although this self-energy contains
off-diagonal terms, it is invariant under the
local SU(2) gauge transformations. 
The normal components are channel-diagonal
\bea
\Sigma_N(\kappa)= \frac{v_{1 {\bf k}}^2}{ \omega - \lambda \tau_3}
+
\frac{v_{2 {\bf k}}^2}{ \omega + \lambda \tau_3},
\eea
but the anomalous terms depend on channel interference:
\bea
\Sigma_A(\kappa) =   -\frac{2\lambda v_{1 {\bf k}}v_{2 {\bf k}}}
{\omega^2 - \lambda^2}
\tau_1,
\eea
and are
directly proportional to the composite 
order parameter $\Lambda$. 
For $\lambda \ne 0$, composite
order induces conventional pairing amongst
the conduction electrons.  We shall later see
that the
independent existence
of the composite order means that 
a finite Meissner
stiffness develops {\sl even when $\lambda=0$, and
conventional pairing is absent}. 

The mean-field free energy per site can be written 
as follows
\bea
F= - T/N_s \sum_{{\bf k} \i \omega_n} 
{\rm Tr \ ln}[ i \omega_n - {\cal H}({\bf k})]
+ 2 \sum_{\Gamma} \frac{v_{\Gamma}^2 }{J_{\Gamma}}.
\eea
The eigenvalues  of the
mean-field Hamiltonian ${\cal H} ({\bf k})$ 
occur in two pairs
$(- \omega_{\bf k \eta},\omega_{\bf k \eta})$, where $\eta  = \pm
$, corresponding to two bands
of quasiparticle excitations. 
We may  rewrite the characteristic determinant
of ${\cal H}({\bf k})$, ${\rm Det} [ \omega-
 {\cal H}({\bf k})] $, in terms of the ${ G}(\kappa)$
\bea
{\rm Det} [ \omega-
{\cal H}({\bf k})]= 
{\rm Det} [ {G}(\kappa)^{-1} ] ( \omega^2 - 
\lambda^2).
\eea
To evaluate the determinant, we write ${G}(\kappa)^{-1}$
in the following form
\bea
{G}(\kappa)^{-1}
=
\bigl(
A - B \tau_3 + C \tau_1\bigr){(\omega^2 - \lambda^2)^{-1}},
\eea
where
\bea
A &=& \omega( \omega^2 - \lambda^2 - v_{{\bf k}}^2),\cr
B&=& \epsilon_{\bf k}( \omega^2 - \lambda^2 )+
\lambda v_{{\bf k} -}^2 ,\cr
C&=& 2\lambda v_{1 \bf k}v_{2 \bf k},\eea
$ v_{{\bf k} }^2 =
 v_{1{\bf k} }^2+  v_{2{\bf k} }^2$, and 
$ v_{{\bf k} -}^2 =
 v_{1{\bf k} }^2\pm  v_{2{\bf k} }^2$.
The eigenvalue equation ${\rm Det}
[\omega - {\cal H}({\bf k})]=0$ then becomes 
\bea 
(A^2-B^2 - C^2)(\omega^2 - \lambda^2)^{-1}   =0.
\eea
Expanding this expression, we obtain
\bea
{\rm Det}[\omega - {\cal H}({\bf k})]=
\omega^4 - 2 \omega^2 \alpha_{\bf k} + \gamma_{\bf k}^2,
\eea
where
\bea
\alpha_{\bf k} = 
v_{{\bf k}}^2 +
{\textstyle \frac{1}{2}}(\epsilon_{\bf k}^2 + \lambda^2), 
\eea
and
\bea
\gamma_{\bf k}
=\sqrt{(\lambda \epsilon_{\bf k}-
v_{{\bf k}-} ^2 )^2 + 
(2 v_{1 \bf k} v_{2\bf k})^2}.
\eea
The eigenvalues of ${\cal H}({\bf k})$ are thus given by
\bea
\omega_{{\bf k}\pm } = \sqrt{ \alpha_{\bf k}
\pm ( \alpha^2_{\bf k}-\gamma ^2_{\bf k})^{1/2}},
\label{spect}
\eea
and 
the mean-field free energy is
\bea
F = - {2 T\over N_s} \sum_{{\bf k}, \eta }
{\rm ln} \biggl[
2 \cosh ( \beta \omega_{{\bf k} \eta}/2)
\biggr]  + 2\sum_{\Gamma=1,2}\frac{(v_{\Gamma})^2 }{J_{\Gamma}}.
\label{free}
\eea
By minimizing the Free energy with respect to
$\lambda$, $v_1$ and  $v_2$, we can now determine the
mean-field phase diagram.  

\section{Phase Diagram}

\begin{figure}[tb]
\twocolumn[\hsize\textwidth\columnwidth\hsize\csname @twocolumnfalse\endcsname
\epsfxsize=6.0truein 
\centerline{\epsfbox{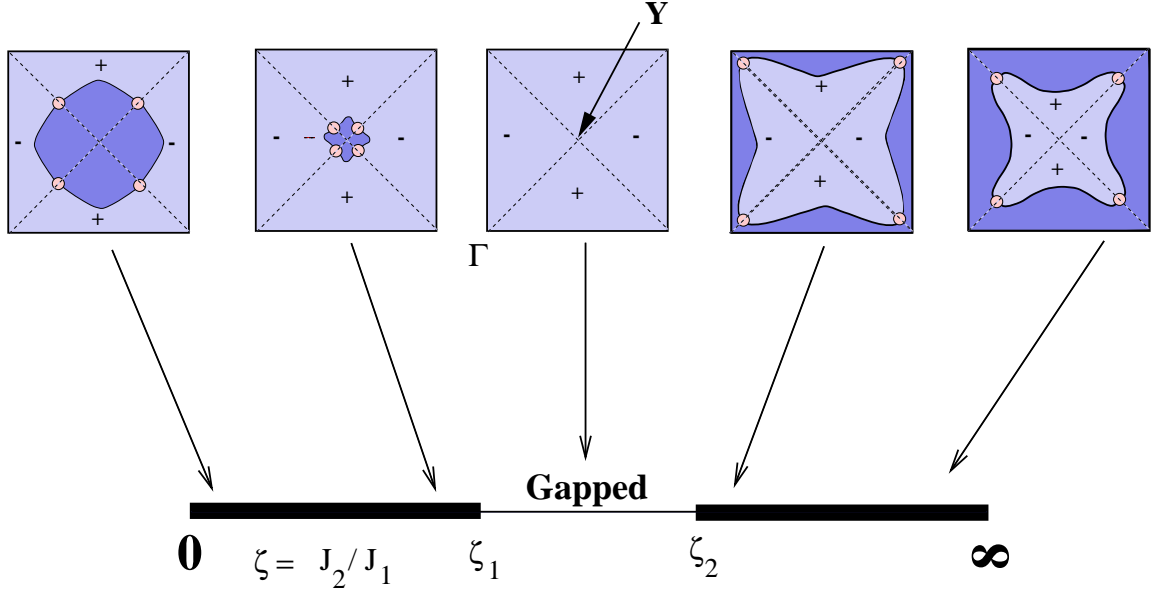}}
\vskip 0.3truein
\protect\caption{Evolution of the gap nodes(open circles) and the underlying
Fermi surface in the composite superconductor, 
as the ratio 
$\zeta = J_2/J_1$ evolves from zero, to infinity. 
In the intermediate region, where $\zeta_1 < J_2/J_1<\zeta_2
$, the underlying Fermi surface collapses around the zone center,
and the nodes annihilate one-another to produce a gapped state.
 }
\label{Fig7}
\vskip 0.2truein]
\end{figure}

We now discuss the mean-field phase diagram.
There are three types of stable mean-field
solution: 

\begin{itemize}
\item{\bf Normal phase.}  $v_1$ or $v_2\neq 0$,
$v_1v_2=0$. At high temperatures, either 
$v_2$ or $v_1$ is finite 
finite, signalling 
a Kondo effect in the
stronger channel.
There are thus {\sl two} types
of normal phase with different Fermi surface geometry, depending on
which channel is the strongest. 
Suppose $v_2=0$,
then the normal state 
spectrum (\ref{spect}) 
attains the simpler form
\bea
\omega_{\bf k \eta} \longrightarrow
E_{\bf k \pm } = \frac{1}{2}\bigl[
(\epsilon_{\bf k} 
+ \lambda) \pm
\sqrt{
({\epsilon_{\bf k} 
- \lambda} )^2 + 4v_{1{\bf k}}^2 
}\bigr],\nonumber
\eea
corresponding to  a  band formed by an admixture
between the conduction electrons, and the
composite f-electrons in channel one.  
This phase describes
a heavy fermion metal. 

\item {\bf Gapless composite paired state.}
$v_1v_2\neq 0$. 
In the generic composite paired ground-state,
the quasiparticle
excitation contains nodes. The condition for gapless excitations
is 
\bea
\gamma_{\bf k}^2 = 0,
\eea
which implies that
\bea
\left.
\begin{array}{l}
(\lambda \epsilon_{\bf k} - v_{{\bf k}-}^2)=0 , \cr
v_{1{\bf k}} v_{2 \bf k}=0.
\end{array}\right\} \label{conditions}
\eea
The first condition defines the locus of points on the underlying
Fermi surface, whilst the  second-condition defines the  nodes of the order
parameter. Gapless quasiparticles
form at the intersection of the order parameter nodes with the
Fermi surface.  This occurs when 
one, or the other channel is dominant and in this case, 
the conduction
propagator (\ref{conduction}) can be written
\bea
G(\kappa)^{-1}= Z_{\bf k}^{-1}[\omega - E_{\bf k}^*\tau_3 - 
\Delta_{\bf k}\tau_1],
\eea
where the gap symmetry 
is determined by the {\sl product} of form factors,
\bea\Delta_{\bf k}=\Delta_o \Phi_{1 {\bf k}}\Phi_{2 {\bf k}},\eea
and 
\bea
Z_{\bf k}^{-1}  = 1 + {v_{{ \bf k}}^2
 \over  \lambda   }
\eea
is a mass-renormalization constant, and 
$E_{\bf k}^* = Z_{\bf k}( \epsilon_{\bf k}
- v_{ {\bf k}-}^2/ \lambda)$, 
$\Delta_o = 2 Z_{\bf k}  v_{1 }
v_{2 }\lambda^{-1}
$
describe the kinetic and pairing contributions
to the quasiparticle energy. 
The heavy electrons in this state are paired, and 
thus correspond to an anisotropic BCS superconductor 
a spectrum
$
\omega_{\bf k} = \sqrt{ (E_{\bf k}^*)^2 +
(\Delta_{\bf k})^2 }
$.

\item{\bf  Gapped composite paired state.} 
$v_1v_2\ne 0$. 
The gapped composite paired phase occurs when
$\zeta= J_2/J_1$ lies between
two critical values, 
\bea \zeta_1 < J_2/J_1<\zeta_2.
\eea
If the weaker  coupling constants is increased
to the point where it is comparable with the stronger channel,
the underlying Fermi surface collapses around the zone center,
causing the 
the nodes to mutually annihilate.  At a a still larger coupling constant,
the nodes reappear at the zone corners.( Fig.\ref{Fig7})
This phenomenon is perhaps easiest to visualize 
when the conduction sea is half-filled. In this case, the normal state is
a
Kondo insulator with  no Fermi surface.\cite{tsunetsugu} The mean-field theory predicts
that when $J_2$ exceeds a critical value, composite pairing can take
place forming a pure composite paired state.
Although the quasiparticle spectrum is gapped, as in a Kondo insulator
\bea
\omega_{\bf k\pm} = \frac{1}{2}\bigl[
\sqrt{
\epsilon_{\bf k} 
^2 + 4v_{+{\bf k}}^2 
}\pm\epsilon_{\bf k} 
\bigr],
\eea
the composite order parameter $\Lambda = 2 v_1v_2 /(J_1J_2)$ is
finite and there is a superconducting response.
When this gapped state is doped, it preserves its gap.

\end{itemize}

In the paired
phase, the mean-field equations are given by the
three conditions
\bea
{\partial F \over \partial \lambda},\   
{\partial F \over \partial v_1} , \ 
{\partial F \over \partial v_2} 
\ 
= 0.
\eea
The first  of these  equations imposes the constraint
$\tilde f\tau_3 f =0$, whereas the second and third 
determine the
magnitude of the Kondo effect in the two channels. 
Written out explicitly, the mean-field equations are
\bea
\frac{1}{N_s}\sum_{\bf k \eta}
\frac{ 
{\rm th} \bigl( \frac{
\omega_{\bf k \eta}}{2T}\bigr)
}
{2 \omega_{\bf k \eta}}\times 
\left\{
 \left( \matrix{\lambda \cr \Phi_{1{\bf k}}^2 \cr  \Phi_{2{\bf k}}^2   } \right)
+ \frac{ \eta }{( \alpha^2_{\bf k}-
\gamma ^2_{\bf k})^{1/2}} {\rm A}
\right\}
=
\left(\matrix{0\cr\frac{2}{J_1}\cr\frac{2}{J_1}}
\right),\nonumber
\eea
\bea
{\rm A}= \left(
\matrix{
\alpha_{\bf k} \lambda-\epsilon_{\bf  k} (\epsilon_{\bf  k} \lambda - v_{{\bf k}-}^2 ) 
\cr
\frac{1}{2}(\epsilon_{\bf k} + \lambda)^2 \Phi_{1{\bf k}}^2
\cr
\frac{1}{2}(\epsilon_{\bf k} - \lambda)^2 \Phi_{2{\bf k}}^2
}
\right).
\eea
Suppose channel one is dominant, then in the normal state
$v_2 =0$, which yields 
two 
equations for the normal state 
\bea
 \sum_ {\bf k,
\eta= \pm}
{\rm th}\bigl( \frac{\beta E_{\bf k \eta}}{2}\bigr)
\biggl[1
-  \frac{ \epsilon_{\bf k} - \lambda }
{E_{\bf k \eta} -
 E_{\bf k -\eta} 
}
\biggr]&=&0,\cr
 \frac{2}{J_1} &=&\chi_{K}(T),
\eea
where
\bea
\chi_K(T)=
\frac{1}{N_s}\sum_{\bf k, \eta= \pm}
{\rm th}\bigl( \frac{\beta E_{\bf k \eta}}{2}\bigr)
\frac{ \Phi_{ 1 \bf k}^2 }{ 
E_{\bf k \eta} -
 E_{\bf k -\eta} 
}
.
\eea
By setting $v_2= 0^+$ in the full set of mean-field equations,
we find
that the transition temperature into the composite paired
state is given by 
\bea
 \frac{2}{J_2} =\chi_{C}(T_c), 
\eea
where
\bea
\chi_C(T) =
\frac{1}{N_s}
\sum_{{\bf k}, \eta = \pm}
{\rm th}\bigl( \frac{\beta E_{\bf k \eta}}{2}\bigr)
\Lambda_{{\bf k}\eta}^2 
\eea
and 
\bea
\Lambda_{{\bf k}\eta}^2 
= 
\Phi_{2 \bf k}^2
\left[
1 + 
\frac{  (\epsilon_{{\bf k}_F} - \lambda) ^2
}
{E_{\bf k -}^2-
E_{\bf k +}^2}
\right]
\eea
is identified as the matrix element $
\Lambda_{{\bf k}\eta}^2 \sim |\langle {{\bf k}_F\uparrow,
-{\bf k}_F\downarrow}\vert \Lambda \vert \Phi\rangle| ^2
$ associated with the action of the composite pair operator
on the Fermi liquid 
ground-state $\vert \Phi\ra$ . In all the above expressions, 
we 
have simplified the algebra using the identities
\bea
E_{\bf k \eta} -
 E_{\bf k -\eta} &=&
\eta \sqrt{
({\epsilon_{\bf k }- \lambda})^2
+
(2 v_{1 \bf k})^2
},\cr
E_{\bf k \eta} +
 E_{\bf k -\eta} &=&
{\epsilon_{\bf k }+ \lambda}.
\eea
We now discuss the detailed phase diagram that
results from the mean-field equations.

\subsection{Instability of the heavy electron metal}
We have argued that the presence of a heavy fermi surface
leads to new zero modes for the transfer of singlets
between different screening channels, so that when a second
channel develops a finite coupling, a composite pair instability
immediately results.
We now describe in detail, 
how this result emerges naturally from our mean-field
theory. 

Let us begin by setting the scale of the Kondo temperature
within this mean-field theory. 
In mean-field
theory, the cross-over into the Fermi-liquid regime is
crudely delineated by a mean-field phase transition.
We use this temperature as a definition of the single
site Kondo temperature. At the ``transition temperature'',
$v_{1}=0^+$, so 
\bea
\chi_K(T_K) &=& \frac{1}{N_s}\sum_{\bf k
}
{\rm th}\bigl( 
\frac{\beta \epsilon_{\bf k }}{2}\bigr)\Phi_{1 \bf k}^2
\frac{
1
}{\epsilon_{\bf k}}\cr
&\approx& 2N(0) \langle \Phi_{1 \bf k}^2\rangle {\rm ln} \left(
{D \over T_K}
\right),
\eea
where we have replaced the momentum sum
by an integral over energy, so that $N(0)$ is the density of
states at the Fermi surface, and 
$\langle \Phi_{1 \bf k}^2\rangle $ denotes a Fermi surface
average of the Form factor. With this definition the mean-field
Kondo temperature takes the form
\bea
T_{K1} \sim D e^{
- 1 / J_1 N(0) \langle \Phi_{1 \bf k}^2 \rangle} .
\eea
This quantity sets the characteristic
size of the mean-field parameters in the normal
state:
\bea
\lambda \sim (V_1)^2 N(0) \sim T_{K1}.
\eea

The composite pair instability of the normal
state is directly related to a divergence in the
fluctuations associated
with the Kondo effect in channel two. 
To see this, we 
expand the 
mean-field expression for the Free
energy to Gaussian order in
${\cal V}_2 =v_{2 x}\tau_x + v_{2 y}\tau_y$,
which gives
\bea
F = F_ o + |v_2|^2 \biggl[\frac{2}{J_2} - \chi_C
\biggr],
\eea
where $v_2= v_{2x} + i v_{2y}$ and $\chi_C$ is given above. 
From this result, we can read off the fluctuations
in the order parameter,
\bea
\langle
\delta v_2  \delta v_2 ^*\rangle
 = \frac{{\rm T }}{\frac{2}{J_2}- \chi_C}.
\eea
But since the composite order parameter is
given by $\delta \Lambda =  2 v_1 \delta v_2 / (J_1J_2)$,
it follows that the composite pair susceptibility
is given by
\bea
\chi_{\Lambda} &= &[\langle \delta \Lambda
\delta \Lambda^* \rangle -
\langle \delta \Lambda
\delta \Lambda^* \rangle_{\chi_C=0}]/T
\cr
&=& 
\left(
{v_1  \over J_1} \right)^ 2
\frac{
\chi_C }{1 - (J_2 \chi_C/2)}.
\eea
The denominator in this expression vanishes
at $T_c$, explicitly confirming that 
the composite pair susceptibility
diverges at the mean-field transition between
the normal, and the paired state.

\begin{figure}[tb]
\epsfxsize=2.6in 
\centerline{\epsfbox{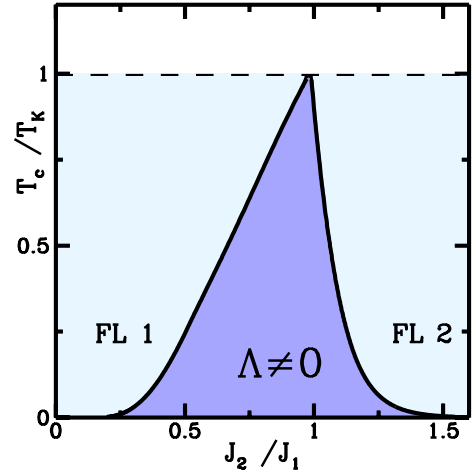}}
\vskip 0.3truein
\protect\caption{Phase diagram
for a two channel Kondo lattice
with ``s'' and ``d-wave'' screening channels.  In
the shaded region, a co-operative Kondo effect in both channels
gives rise to composite pairing and a quasiparticle
gap with  $s\times d = d$-wave symmetry. 
 }
\label{Fig4}
\end{figure}

To gain some insight into the composite pair
instability, we divide the bare composite
susceptibility, 
$\chi_C$ into a ``high'' and  a ``low''
energy component
\bea
\chi_C = \sum_{|E_{{\bf k}\eta}|> T_{K1}} +
 \sum_{|E_{{\bf k}\eta}|<  T_{K1}}\left\{ \dots \right\}
=\chi_h + \chi_l,
\eea
where the former  describes the local Kondo effect
in the weaker channel, the latter, 
the channel interference taking place on the heavy Fermi surface.
At energies $|E_{{\bf k}\eta}|>> T_{K1}$, 
\bea
\sum_{\eta}{\rm th}\bigl( \frac{\beta E_{\bf k \eta}}{2}\bigr)
\frac{\Lambda_{\bf k \eta}^2}
{2  E_{\bf k\eta}}
\rarrow 
\frac{
1
}{\vert \epsilon_{\bf k}\vert }\Phi_{2 \bf k}^2,
\eea
Suppose the heavy Fermi surface lies in the lower band, 
on the Fermi surface,
\bea
\epsilon_{\bf k_F} &=& \frac{v_{1 \bf
k}^2 }{\lambda} \cr E_{{\bf k_F}-}^2
-E_{{\bf k_F}+}^2 &=& - (\lambda^2 + v_{1
\bf k}^2 )/\lambda
\eea
so that 
\bea
\Lambda_{{\bf k}_F \eta}^2
=
\frac{4 \lambda^2 v_{1 }^2
(\Phi_{1 \bf k_F}\Phi_{2 \bf k_F})^2
}{(\lambda^2
+ v_{1 \bf k_F}^2)^2}.
\eea
When we replace the momentum sums by energy integrals, we must
remember that the density of quasiparticle states is enhanced
by a factor
\bea
N_{\bf k_F}^*(0) =
N(0) 
\left. \frac
{d \epsilon_{\bf k}}
{d E_{{\bf k}-}}
\right|_{{\bf k}={\bf k}_F}
= \frac{(\lambda^2
+ v_{1 \bf k}^2)}{\lambda^2}N(0).
\eea
The energy scale separating the two regimes is the Kondo temperature
for channel one, $T_{K1}$. With these results, approximate
expressions for the high and low energy contributions to the
composite susceptibility are
\bea
\frac{\chi_h}{2N(0)} 
&=& 
\int_{T_{K1}}^D
\frac{d \epsilon }{\epsilon} \langle
\Phi_{2 \bf k}^2\rangle ,
\cr
\frac{\chi_l}{2N(0)} 
&=& 
2\int_T^{T_{K1}}
 \frac{d E}{E} \left\langle 
\frac{ (v_{1 }\Phi_{1 \bf k}\Phi_{2 \bf
k})^2}{(\lambda^2
+ v_{1 \bf k}^2)}
\right\rangle\cr
&\approx&
2
\int_T^{T_{K1}}
 \frac{d E}{E} \langle 
\Phi_{2 \bf k}^2\rangle.
\eea
The expression for $\chi_l$ was simplified
by noting that the dominant contribution to the second
term occurs in the regions far from the node
in the order parameter,  where
$v_{1\bf k}>> \lambda$.  The sum of the two expressions then yields
\bea
\chi_C &\approx & 2 N(0)\biggl[{\rm ln}\bigl(
{D \over T_{K1}}\bigr)
+ 2{\rm ln} \bigl(
{T_{K1} \over T}\bigr)
\biggr]\langle \Phi_{2 \bf
k}^2\rangle.
\eea
The Gaussian coefficient of $v_2$ in the
Free energy is thus given by
\bea
\frac{2 }{J_2} &-& \chi_C \cr
&=&
 2N(0)\langle \Phi_{2 \bf
k}^2\rangle
\left[ \frac{1}{g_2}-
{\rm ln}\bigl(
{D \over T_{K1}}\bigr)
- 2{\rm ln} \bigl(
{T_{K1} \over T}\bigr)
\right],
\eea
where 
$
g_2= N(0)J_2 \langle \Phi_{2 \bf
k}^2\rangle$ is the dimensionless Kondo coupling constant
for channel two. 
The first logarithm in this expression describes the
renormalization of the coupling constant in channel two
down to the energy scale $T_{K1}$:
\bea
 \frac{1}{g_2(T_{K1})} =  \frac{1}{g_2}-
{\rm ln}\bigl(
{D \over T_{K1}}\bigr);
\eea
the second logarithm describes the subsequent renormalization
of $g_2$ at temperatures below $T_{K1}$.
In a two-channel single impurity model, the second logarithm
would be entirely
entirely absent because the Kondo effect in channel one
cuts off any further renormalization in channel two. 
Here we see that the constructive interference between the two
channels in the lattice 
actually 
over-compensates for the Kondo effect in channel one,
producing a logarithmic renormalization at low temperatures
which is twice as large as at high temperatures. The co-operative
Kondo effect thus develops at a temperature which is higher
than the Kondo temperature for an isolated channel two. 
We may rewrite the Gaussian coefficient as 
\bea
\frac{2 }{J_2} - \chi_C &=&\frac{2 }{J_2} - 
 2 N(0)\biggl[{\rm ln}\bigl(
{D \over T}\bigr)
+ {\rm ln} \bigl(
{T_{K1} \over T}\bigr)
\biggr]\langle \Phi_{2 \bf
k}^2\rangle\cr
&=& - 
 2 N(0)\biggl[{\rm ln}\bigl(
{T_{K2}\over T}\bigr)
+  {\rm ln}\bigl(
{T_{K1} \over T}\bigr)
\biggr]\langle \Phi_{2 \bf
k}^2\rangle\cr
&=& 
 4 N(0)\langle \Phi_{2 \bf
k}^2\rangle{\rm ln}\bigl(
{T\over \sqrt{T_{K1}T_{K2}} }\bigr),
\eea
where we have used the definition
\bea T_{K2} = D e^{ -1 / (N(0) J_2 
\langle \Phi_{2 \bf k}^2\rangle) },\eea to absorb
the coupling constant
$J_2$. 
In this rough approximation, the composite
pairing instability occurs at a temperature
\bea
T_c \sim \sqrt{T_{K1}T_{K2}}.
\eea
When $J_2/J_1<<1$, this same scale sets the size of $\Delta_o$.
Fig. \ref{Fig4} illustrates the  phase
diagram calculated numerically 
for a two channel Kondo lattice
with ``s'' and ``d-wave'' screening channels:
\bea
\begin{array}{rcll}
\Phi_{1\bf k}&=&1,\qquad &\hbox{(s-channel)}\cr
\Phi_{2\bf k}&=&(\cos k_x - \cos k_y), \qquad &\hbox{($d_{x^2 - y^2}$-channel)}.
\end{array}
\label{sd}
\eea
as shown in Fig. \ref{fig2a}. 
As expected, the composite pair instability
occurs at the  highest temperature when
the two channels are most evenly matched. 

\subsection{Composite pair instability of the
Kondo insulator}

Since composite
pairing is an intrinsically local process,
the presence of a heavy fermi surface is not
a  necessary requirement for the formation of
the paired state, but in its absence,
the instability
requires the  second-channel coupling
constant to exceeds a critical value. 

\begin{figure}[tb]
\epsfysize=4.0in
\centerline{\epsfbox{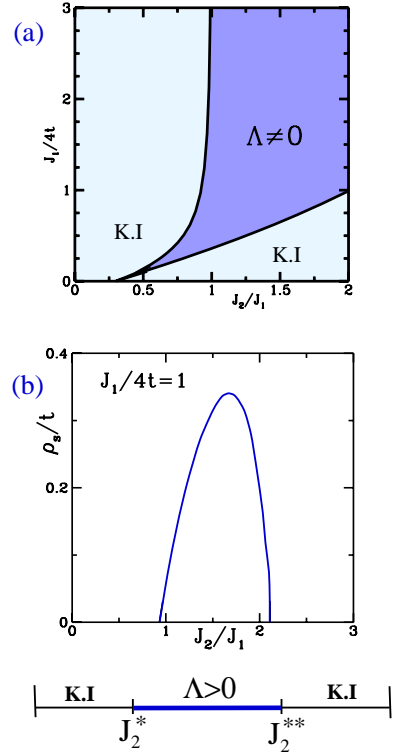}}
\vskip 0.3truein
\protect\caption{(a) Phase diagram for two-channel Kondo insulator.
``K.I ''  denotes  the ``Kondo insulating'' phases.
In the intermediate gapless
phase both channels participate coherently in the composite pairing
process.
(b)
Showing phase stiffness in the composite  paired
phase for the case $J_1/4t =1$.}
\label{Fig5}
\end{figure}

This is precisely what happens 
where the conduction band is half filled, for 
in this case, the normal state of the Kondo
lattice is a ``Kondo insulator'', with no Fermi
surface and a 
gap to charge excitations.  The Kondo insulator
is particle-hole symmetric, so in the mean-field
theory, $\lambda=0$, so that the excitation
spectrum simplifies to the following
form
\bea
\omega_{{\bf k}\pm}= \sqrt{\bigl(
{\epsilon_{\bf k} \over 2}
\bigr)+ v_{{\bf k}}^2} \pm
{\epsilon_{\bf k} \over 2}.
\eea
By minimizing the mean-field Free-energy with
respect to variations in $v_1$ and $v_2$, and setting
$T=0$, we 
obtain two mean-field equations for the ground-state
\bea
{ 1\over J_1} &=&
\int \frac{d^dk}{(2 \pi)^d}
\frac{
\Phi_{{\bf k} 1}^2 }
{
\sqrt{
{\epsilon_{\bf k}}^2
+ (2 v_{{\bf k}})^2}},
\cr
{1 \over J_2} &=&
\int \frac{d^dk}{(2 \pi)^d}
{\Phi_{{\bf k} 2}^2 \over
\sqrt{
{\epsilon_{\bf k}}^2
+ (2 v_{{\bf k}})^2}},\label{instabil1}
\eea
(where $d$ is the dimensionality).
This composite paired phase will only be
stable within a range of $J_2$, $J_2^*< J_2 < J_2^{**}$.
By setting $v_2=0^+$, we obtain two parametric
equations for the second-order phase boundary between the Kondo
insulator and the composite paired state:
\bea
{ 1\over J_1} &=&
\int \frac{d^dk}{(2 \pi)^d}
\frac{
\Phi_{{\bf k} 1}^2 }
{
\sqrt{
{\epsilon_{\bf k}}^2
+ ( 2 v_1 \Phi_{1 {\bf k}})^2}},
\cr
{1 \over J_2^*} &=&
\int \frac{d^dk}{(2 \pi)^d}
{\Phi_{{\bf k} 2}^2 \over
\sqrt{
{\epsilon_{\bf k}}^2
+ ( 2 v_1 \Phi_{1 {\bf k}})^2
}}.\label{instabil3}
\eea
Beyond the critical value $J_2^{**}$, the Kondo effect
is no longer operative in channel two. By setting
$v_1 = 0^+$, we obtain two parametric equations for
the second phase boundary. 
\bea
{ 1\over J_1} &=&
\int \frac{d^dk}{(2 \pi)^d}
\frac{
\Phi_{{\bf k} 1}^2 }
{
\sqrt{
{\epsilon_{\bf k}}^2
+ ( 2 v_2 \Phi_{2 {\bf k}})^2}},
\cr
{1 \over J_2^{**}} &=&
\int \frac{d^dk}{(2 \pi)^d}
{\Phi_{{\bf k} 2}^2 \over
\sqrt{
{\epsilon_{\bf k}}^2
+ ( 2 v_2 \Phi_{2 {\bf k}})^2
}}.\label{instabil4}
\eea
We have calculated the phase diagram for
a two-dimensional 
Kondo insulator in two dimensions, with
dispersion
$\epsilon_{\bf k } = -2t (c_x +c_y)$ and
a Kondo coupling in the s- and d-channel, as
shown in equation (\ref{sd}).
When $J_2< J_2^*$, the Kondo effect in the
s-channel leads to Kondo insulator.
By contrast, when $J_1<< J_2$, a 
``nodal semi-metal''  forms in channel two,
with a gap which vanishes along the
d-wave nodes of the  form-factor $\Phi_{2 \bf k}
$.
Fig. \ref{Fig5} shows the phase diagram
obtained from equations (\ref{instabil3}) and (\ref{instabil4}).
Notice that 
the range of $J_2$ over which the composite paired
state is stable, is negligible for small $J_1$ and $J_2$, 
but grows substantially as both $J_1$
becomes large compared with $t$.

This  composite paired state 
is interesting, for its quasiparticle
spectrum is essentially identical to the Kondo
insulator, and furthermore, since $\lambda=0$,
there is no
anomalous component to the conduction electron
self energy: there are no paired conduction electrons
in the ground-state.
Yet despite these similarities, 
the presence of composite order
\bea
\Lambda(x) = {2
 \over J_1  J_2}v_1 v_2 e^{- i \phi(x)},
\eea
means that this state is a superconductor, with a finite
charge susceptibility. The calculation of the charge susceptibility
needs to be carried out subject to the constraint.  Expanding the 
Free energy energy to quadratic order in changes in $\lambda$ and the
chemical potential, $\mu$ we have
\bea
F= F_o - \frac{1}{2}\bigl[ \chi_{\mu \mu} (\delta \mu)^2 +  
2\chi_{\mu \lambda}\delta \mu \delta \lambda + \chi_{\lambda \lambda}
(\delta \lambda )^2 \bigr].
\eea
The constraint $\partial F/ \partial {\lambda} =0$ implies
that
\bea
\delta \lambda = - \frac{\chi_{\mu \lambda}}{\chi_{\lambda \lambda}}
\delta \mu,
\eea
so that we may write
\bea
F= F_o - \frac{\chi_C}{2} (\delta \mu)^2,
\eea
where
\bea
\chi_C = \chi_{\mu \mu} - \frac{ \chi_{\mu \lambda }^2
}{\chi_{\lambda \lambda}}.
\eea
A rather laborious calculation (Appendix D) gives
\bea
\chi_{\mu \mu} &=& \sum _{\bf k} \frac{4v_{{\bf k}}^2 }{[(\epsilon_{\bf k})^2
+ 4v_{{\bf k}}^2]^{3/2}},\cr
\chi_{\mu \lambda} &=& \sum _{\bf k} \frac{4v_{{\bf k}-}^2 }{[(\epsilon_{\bf k})^2
+ 4v_{{\bf k}}^2]^{3/2}},\cr
\chi_{\lambda \lambda} &=& \chi_{\mu \mu} + \chi_b,
\eea
where 
\bea
\chi_b = \sum _{\bf k} \frac{\epsilon_{\bf k}^2 }{[(\epsilon_{\bf k})^2
+ 4v_{{\bf k}}^2]^{3/2}}\left(\frac{ 2 v_{1{\bf k}}v_{2{\bf k}}}
{v_{{\bf k}}^2}
\right)^2({5} + \frac{\epsilon_{\bf k}^2 }{v_{\bf k}^2}).
\eea
Since $|\chi_{\mu \lambda}| \le \chi_{\mu \mu}$ and
$\chi_{\lambda \lambda } \ge \chi_{\mu \mu}$,
$\chi_C$ is positive, provided both $v_1$ and $v_2$ are finite. 
In the next section we also 
confirm that the composite paired state has a 
finite superfluid phase stiffness, given by 
\bea
\rho_s = {1 \over d}\sum_{\bf k} {(v_1v_2)^2 \over 
v_{\bf k}^2}
{
(\Phi_{1\bf k}\pmb{$\nabla$}\Phi_{2\bf k}-\Phi_{2\bf
k}\pmb{$\nabla$}\Phi_{1\bf k})^2\over
[\epsilon_{\bf k}^2 + (2 v_{\bf k+})^2]^{\frac{1}{2}}},
\eea
in the ground-state. 

\subsection{Gapped Composite Paired State}

In our simple two dimensional example, $v_{1\bf k}$ is never
zero, so the condition (\ref{conditions})
for gapless behavior become 
\bea
\left. \begin{array}{l}
\lambda = \frac{v_{1{\bf k}}^2}{\epsilon_{\bf k}  }, \cr
v_{2 \bf k}=0.
\end{array}\right\}
\eea
The modulus  $\vert(v_{1{\bf k}}^2/\epsilon_{\bf k}  )\vert $ is 
smallest at the band-edges, where
where $\vert \epsilon_{\bk }\vert = 
W_{\pm} = 4t \mp \mu$. So if 
\bea
-\left(\frac{v_{1{\bf k}}^2}{W_-}\right)<
\lambda<
\left(\frac{v_{1{\bf k}}^2}{W_+}\right),
\eea
the paired state
becomes gapless.  
Although a small residual Cooper pair
density is present, 
the state is essentially 
a pure condensate of composite pairs.
To understand what happens at the critical values of $\zeta$,
it is instructive to examine the underlying Fermi surface, defined
by the locus of points 
\bea
\epsilon_{\bf k} = \frac{v_{1{\bf k}}^2 - v_{2{\bf k}}^2}{\lambda}.
\eea
As one approaches the critical value $\zeta= \zeta_1$, this Fermi surface
collapses around the zone-center, forcing the nodes
to mutally annihilate.
In going from the region $J_2/J_1<<1$ to 
the region $J_2/J_1>>1$, $\lambda$ changes sign, and so that
the system must always pass through this region of gapless
composite pairing. 
In the center of this region, where  $\lambda=0$, the state
is a pure composite paired state.
Once $\zeta$ exceeds the value $\zeta_2$, 
the nodes are reborn at the zone corners and 
the Fermi-surface re-appears along the zone-boundary (Fig. \ref{Fig7}.)

To illustrate these conclusions, we have used the mean-field
equations to calculate numerically
the two lines where the gap vanishes.  In Fig. \ref{Fig5q}, we summarize
the results of these calculations, in a diagram where we have
kept ${\hbox{max}}(J_1,J_2)=4t$, and varied the ratio $J_2/J_1$. 

\section{Superfluid Density of the composite paired state}

To confirm that the composite paired state is superconducting,
we need to compute the superfluid density $\rho_s$.  In the London gauge
$\nabla\cdot  {\bf A} =0$, the supercurrent is given by
\bea
{\bf j}_s = - Q {\bf A},
\eea
where 
\bea
Q_{ab}= e^2 [ \rho_{s}]_{ab}= {\partial^2 F \over {\partial
{\bf A}_a}{\partial {\bf A}_b}}.
\eea
In the presence of an electro-magnetic field, the electron kinetic
energy {\sl and}  form factors acquire a dependence on the
vector potential. Using a Nambu notation, we may write
\bea
\epsilon_{\bf k} &\rightarrow & \epsilon_{{\bf k}- e {\bf A}\tau_3},\cr
\Phi_{\Gamma \bf k} &\rightarrow & \Phi_{\Gamma {\bf k}- e {\bf A}\tau_3},
\eea
so that the hybridization acquires the form 
\bea
{\cal V}_{{\bf k}}^{\bf A} =  i v_1 \Phi_{1 {\bf k}- e {\bf A} \tau_3}
+ v_2 \tau_2\Phi_{2 {\bf k}- e {\bf A} \tau_3}..
\eea
The appearance of the vector potential in the form factor reflects
the fact that the hop and flip motion of electrons around a
local moment leads to current flow. 
In the London-gauge ($\nabla \cdot {\bf A}=0$) we may calculate
by compute the second-derivative
of the Free energy at fixed values of $v_1$, $v_2$ and $\lambda$,\cite{agd}
so the the only important part of the Free energy is the electronic component:
\bea
F_e = - T \sum_{\kappa} {\rm Tr} {\rm ln} [i \omega_n - {\cal H}_{\bf A}({\bf k})].
\eea
To second-order in the vector potential, we may write
\bea
{\cal H}_{\bf A}({\bf k})= {\cal H}({\bf k})- {\cal J}^a A_a +
\frac{1}{2}A_aA_b\nabla^2_{ab}{\cal H}({\bf k})+ O(A^3),
\eea
so expanding the Free energy to second-order in ${\bf A}$, we obtain
\bea
F&=&F_o + {\textstyle\frac{1}{2}}Q_{ab}A_aA_b,
\cr
Q^{ab} &=&  T 
 \sum_{\kappa} \biggl\{
{\rm Tr}  [{\cal G}_{\kappa}{\cal J}^{a}_{\kappa}
{\cal G}_{\kappa}{\cal J}^{b}_{\kappa}] +
 [{\cal G}_{\kappa}\nabla^2_{ab}{\cal H}_{\kappa}
] 
\biggr\},\label{twoterms}
\eea
where 
\bea
{\cal G}_{\kappa} = [ i \omega_n - {\cal H}({\bf k})]^{-1}\equiv
\left[\matrix{
G &, G_{cf}\cr G_{fc} & G_{ff}}
\right],
\eea
is the matrix propagator and 
\bea
{\cal J}_{\kappa} = - \nabla_{A} {\cal H}_A({\bf k})=
e \left[\matrix{ \pmb{$\nabla$}_{{\bf k}}
\epsilon_{\bf k} & \tau_3\pmb{$\nabla$}_{{\bf k}}{\cal V}\dg _{{\bf k}} \cr
\pmb{$\nabla$}_{{\bf k}} {\cal V}_{{\bf k}} \tau_3 &0
}\right]
\eea
is the current operator. 
The first and second 
terms in (\ref{twoterms}) correspond to the paramagnetic
and diamagnetic components of the stiffness, as in a conventional
superconductor. 
Notice however that the current contains
anomalous off-diagonal contributions that do not commute
with the charge operator $\tau_3$. Unlike a pure 
BCS superconductor, here the presence of composite pairs
affects the current operator. 

The diamagnetic contribution to 
$Q^{ab}$ can integrated by parts, to obtain
\bea
 \sum_{\kappa} 
{\rm Tr}  [{\cal G}_{\kappa}\nabla^2_{ab}{\cal H}_{\kappa}]
=
-  \sum_{\kappa} 
{\rm Tr}  [{\cal G}_{\kappa}{\rm j}^{a}_{\kappa}
{\cal G}_{\kappa}{\rm j}^{b}_{\kappa}], \eea
where
\bea
{\bf \rm j}_{\kappa} =  e {\bf \nabla}_{\bf k} {\cal H}({\bf k})
=e \left[\matrix{
{\bf \nabla}_{\bf k}\epsilon_{\bf k}\tau_3 & {\bf \nabla}_{\bf k}{\cal V}\dg _{\bf k} \cr
 {\bf \nabla}_{\bf k}{\cal V}_{\bf k} &0
}\right],
\eea
so that the full expression for the superfluid stiffness tensor
is
\bea
Q_{ab} = T 
 \sum_{\kappa} \biggl\{
{\rm Tr}  [{\cal G}_{\kappa}{\cal J}^{a}_{\kappa}
{\cal G}_{\kappa}{\cal J}^{b}_{\kappa}] -
{\rm Tr}  [{\cal G}_{\kappa}{\rm j}^{a}_{\kappa}
{\cal G}_{\kappa}{\rm j}^{b}_{\kappa}]
\biggr\}.
\eea
To gain some insight into this equation,  it
is instructive to evaluate the stiffness for the case of
pure composite pairing at half-filling. 
In this case, the conduction electron
propagator commutes with the charge operator $\tau_3$, 
so contributions to the
stiffness which involve the conduction electron component of
${\cal G}$ identically vanish. For example, the cross-term
between $\pmb{$\nabla$}\epsilon_{\bf k}$ and $\pmb{$\nabla$}{\cal
V}$ is 
\bea
 &&
{\rm Tr}  \bigl[G_{cc}\nabla_a\epsilon G_{cf}\nabla _b{\cal V}\tau_3
- G_{cc}\nabla_a\epsilon\tau_3 G_{cf}\nabla _b{\cal V}\bigr]
 = 0.\label{cross}
\eea
The only surviving terms represent the composite pair stiffness,
which can be written in the form
\bea
Q_{ab}^C =  \frac{\rm T}{2}{\rm Tr} \bigl[
[G_{cf} \nabla_a {\cal V},\tau_3] 
[G_{cf} \nabla_b {\cal V},\tau_3] +{\rm H. c.}\bigr].
\eea
From the Dyson equation, 
\bea
{\cal G}( \kappa) 
={\cal G}_o( \kappa) 
+{\cal G}
(\kappa)  
\left[
\matrix{
0
& {\cal V} \dg 
\cr
{\cal V} 
& 0  
} \right]{\cal G}_o( \kappa), 
\eea
where ${\cal G}_o$ is the propagator in the absence of any hybridization,
it follows that 
\bea
G_{cf}(\kappa) =  G(\kappa)
{\cal V}\dg\frac{1}{\omega - \lambda \tau_3},
\eea
so that at half-filling ($\lambda =0$), 
\bea
G_{cf}(\kappa) = \frac{1}{\omega ( \omega - \epsilon_{\bf k} \tau_3)
- v_{{\bf k}}^2 } {\cal V}\dg,
\eea
and hence
\bea
[ G_{cf}\pmb{$\nabla$}{\cal V}, \tau_3] = 
\frac{1}{\omega ( \omega - \epsilon_{\bf k} \tau_3)
- v_{{\bf k}}^2 }[ {\cal V}\dg \pmb{$\nabla$}{\cal V}, \tau_3].
\eea
Evaluating the commutator 
\bea
[ {\cal V}\dg \pmb{$\nabla$}{\cal V}, \tau_3]
 = 2( v_{1 {\bf k }} \pmb{$\nabla$}v_{2{\bf k } }
- 
v_{2 {\bf k }} \pmb{$\nabla$}v_{1{\bf k }} ) \tau_1,
\eea
we may then write
\bea
Q^C = \frac{4 e^2  T }{d} \sum_{\kappa}
( v_{1 {\bf k }} \pmb{$\nabla$}v_{2{\bf k } }
- 
v_{2 {\bf k }} \pmb{$\nabla$}v_{1{\bf k }} )^2 
{\rm Tr} 
\bigl[ G(\kappa)^2\bigr],
\eea
where for convenience, we have assumed an isotropic stiffness
$Q_{ab}^C= Q^C \delta_{ab}$. 
Taking the zero-temperature limit, replacing $i \omega_n \rarrow
i \omega$ and 
\bea
{\rm T} \sum_{i \omega_n}
\rarrow \int _{\- \infty} ^{\infty}
\frac{ d \omega }{2 \pi},
\eea
then yields
\bea
e ^2 \rho^C_s &= &
\frac{8 e^2 }{d} \sum_{\bf k}
( v_{1 {\bf k }} \pmb{$\nabla$}v_{2{\bf k } }
- 
v_{2 {\bf k }} \pmb{$\nabla$}v_{1{\bf k }} )^2\cr 
&\times &\int_{- \infty}^{\infty}
\frac{ d \omega }{2 \pi}
\frac{1}{( \omega^2 + v_{{\bf k}}^2 )^2 + \omega^2 \epsilon_{\bf
k}^2 }.\eea
Carrying out the final frequency integral yields the following
expression for the composite stiffness 
\bea
e^2 \rho_s^C = {e^2  \over d}\sum_{\bf k} \left({v_1v_2 \over 
v_{\bf k+}}
\right)^2 {
(\Phi_{1\bf k}\pmb{$\nabla$}\Phi_{2\bf k}-\Phi_{2\bf
k}\pmb{$\nabla$}\Phi_{1\bf k})^2\over
[\epsilon_{\bf k}^2 + (2 v_{\bf k+})^2]^{\frac{1}{2}}}.
\eea
This result is quite fascinating, because it confirms that the composite
pair condensate has a phase stiffness, even in the absence
of an underlying Fermi surface. In a conventional superconductor,
the scale of the stiffness is determined by the
Fermi energy $e^2 \rho_s \sim e^2 \epsilon_F/a^2$. Here 
the size of the stiffness 
\bea
e^2 \rho_s^C \sim  e^2 N(0)\left\langle
\left({v_1v_2 \over 
v_{\bf k+}}
\right)^2 \right\rangle
\sim e^2 T_c/a^2
\eea
is determined by the condensation energy.
This is a classic  example of ``local pair''
condensation.

Away from half filling, the superfluid stiffness contains
contributions associated with both the paired conduction electrons,
and the composite pairs. To make a clean division of the stiffness
into two components, one needs to worry about the various cross-terms
that appear in the stiffness such as (\ref{cross}), which do not
obviously vanish away from half filling.  
However, a key observation is that
these terms are odd functions of $\lambda$, 
so they are guaranteed guaranteed to vanish provided that the physics is
particle-hole symmetric about $\lambda=0$. With this proviso, 
we can divide the superfluid
stiffness into two terms
\bea
{Q}= {Q}^{BCS} + {Q}^{C},
\eea
where 
\bea
Q^{BCS}_{ab}&=& \frac{T}{2}\sum_{\kappa}
\nabla_a \epsilon_{\bf k}
\nabla_b \epsilon_{\bf k}
{\rm Tr}\biggl[ [ i G( \kappa), \tau_3]^2\biggr],\cr
Q_{ab}^C &=&  \frac{\rm T}{2}{\rm Tr} \bigl[
[G_{cf} \nabla_a {\cal V},\tau_3] 
[G_{cf} \nabla_b {\cal V},\tau_3] +{\rm H. c.}\bigr].
\eea
are the BCS and composite pair contributions
to the stiffness.

\section{Discussion}

In this discussion, we should like to address the results of
this paper on two fronts:
\begin{itemize}

\item Alternative theoretical approaches to test and confirm
the presence of a  co-operative Kondo effect.

\item Applications to experiments and the theory of real
heavy fermion systems. 

\end{itemize}

\subsection{Alternative Theoretical Approaches}

The main effort of this paper has been to establish the
link between co-operative channel interference in the Kondo lattice
and composite pairing.  Although our key result- the
relation between the gauge invariant interference term
and composite order
\bea
{\cal V}\dg_2{\cal V}_1 = - \frac{J_1J_2}{2}
\left[
\matrix{
F\dg &  \Lambda\cr
- \Lambda \dg & F\dg 
}
\right],
\eea
does not depend on approximations, the notion that
this interference term can develop long-range order relies
on various mean-field approximations.  
There are however reasons to be confident in the mean-field solution.
First, there are no obvious competing instabilities, such as
magnetism. By turning on a $J_2$, one does not drive the
system closer to the antiferromagnetic instability, but rather,
simply activates another source of Kondo screening. This should
be contrasted with the situation in the alternative 
model of spin-fluctuation mediated pairing.
Here, 
to attain a transition temperature that is comparable
with the heavy fermion band-width places the model close
to an antiferromagnetic instability, where the competing
effects of magnetism make the mean-field theory potentially
unreliable.

Nevertheless, the use of mean-field methods
inevitably raises questions
about our work which motivates us to seek
alternative methods to verify the key results.
It may be possible
to precisely verify our results  in both finite size calculations,
and in the exactly solvable limit of 
infinite dimensions.
Finite size studies on our model may be facilitated by treating
the model as a ladder compound and by using the strong-coupling limit
so as to completely eliminate the possibility of antiferromagnetic
instabilities. 

To formulate our model in a form that is tractable to 
an exact infinite dimensional study, rather than using explicit form
factors in the Kondo interaction, 
it is better to start with
a channel conserving two-channel Kondo lattice, to which 
a term which destroys channel conservation is then added
to activate the channel interference.  
Suppose one
starts out with a  two-channel Kondo 
model, with perfect channel conservation:
\bea
H^{C} =\overbrace{
\sum_{\bf k \Gamma \sigma} 
\epsilon_{\bf k}
c{^{\dag}}_{\Gamma{\bf k}\sigma} 
c_{\Gamma{\bf k}\sigma}}^{H_o} +
\sum_{\Gamma j} J_{\Gamma}
c{^{\dag}}_{\Gamma j}
\pmb{$\sigma$} c_{\Gamma j} \cdot {\bf S_j}.
\eea
where 
\bea
c{^{\dag}}_{\Gamma j\sigma } ={N_s}^{-\frac{1}{2}}
\sum_{\bf k} 
c{^{\dag}}_{\Gamma {\bf k}\sigma} e^{-i {\bf k}
\cdot {\bf R}_j}, 
\eea
$ (\Gamma = 1, \ 2)$ and
\bea
\epsilon&=& -2t \sum_{l=1,d}(\cos k_l) - \mu.
\eea
This model has a perfect $U(1)$  channel symmetry. 
Suppose one now adds a  term to the hopping 
$
H \rarrow H^{C} + H'$, where
\bea
H' = - 2 \Delta  \sum_{l=1,d} (-1)^l\cos {k_l}
(c \dg_{{\bf k} 1 \sigma}
c _{{\bf k} 2 \sigma}+
c \dg_{{\bf k} 2 \sigma}
c _{{\bf k} 1 \sigma} ),
\eea
is a hopping term with ``d-wave'' symmetry that mixes the different
channels, breaking the $U(1)$ channel conservation symmetry. 
This contains no 
no non-local interactions, and is thus ideal for a large
d-treatment. It is easy to see that in
the limit of large $\Delta$, it is equivalent 
to a one band model, with two orthogonal form factors
\bea
\Phi_{1{\bf k}} &=& 1,\cr
\Phi_{2{\bf k}} &=& {\rm sgn}( {\textstyle \sum_l (-1)^l \cos k_l}).
\eea
To see this, note that when $\Delta$ is non-zero, the band
splits into two components with energies 
\bea
E_{{\bf k} \pm } = \epsilon_{\bf k} \pm 2 \Delta \left| \sum_{l=1,3, \dots} (-1)^l 
\cos {k_l}\right |
\eea
where 
\bea
H_o + H'& =& \sum_{\bf k \sigma} (E_{\bf k+ }
a
 \dg_{\bf k \sigma} a_{\bf k \sigma}+ E_{\bf k- } b\dg_{{\bf k} \sigma } b_{{\bf k} \sigma }),
\cr
a_{{\bf k} \sigma} &=& \frac{1}{\sqrt{2}}
(c_{\bf k 1\sigma}+ \Phi_{2\bf k} c_{\bf k 2\sigma}),\cr
b_{ {\bf k} \sigma} &=& \frac{1}{\sqrt{2}}
(c_{{\bf k} 1\sigma}- \Phi_{2\bf k} c_{{\bf k} 2\sigma}).
\eea
At large $\Delta$, one can project out all terms  involving the upper
band by re-writing the Hamiltonian in terms of
the $a$ and $b$ creation operators, then all terms involving
the upper band creation or annihilation operators
$a_{\bf k\sigma}$ or $a\dg_{\bf k\sigma}$.  When we do this,
the interaction becomes  
\bea
H_I &=& \frac{1}{2 N_s}\sum_{{\bf k}, {\bf k'}} 
J_{{\bf k},{\bf k'}}
 b\dg_{\bf k}\pmb{$\sigma$}b_{\bf k} \cdot S_j
e^{ i ({\bf k} - {\bf k'}) \cdot {\bf x}_j},\cr
J_{{\bf k},{\bf k'}}&=& ( 
J_1 \Phi_{1 {\bf k}}\Phi_{1 {\bf k'}}+
J_2 \Phi_{2 {\bf k}}\Phi_{2 {\bf k'}}
),
\eea
corresponding to a two-channel Kondo model with two orthogonal
matrix elements.  

The mean-field theory for this model predicts that 
that as soon as $\Delta$ becomes finite,
channel interference will drive the system
into a composite paired state, even when $J_1>> J_2$. 
This phenomenon should extend all the way out to infinite dimensions,
where a precise dynamical mean-field theory treatment of the model
becomes possible. 

\subsection{Possible consequences of co-operative
Kondo behavior for Heavy Fermion compounds }

We conclude with a brief discussion of the general
consequences of co-operative channel interference in heavy fermion systems.
Our paper has focussed on the superconducting aspects of this problem.
Here we should like to put the problem in a more general perspective.

Assuming it becomes possible to verify the theoretical soundness of the
co-operative Kondo effect, how could the theory be tested experimentally?

\begin{itemize}

\item Our model suggests a rather intimate relation between
the local quantum chemistry of the heavy fermion ion, and the gap symmetry of the order parameter.   In heavy fermion compounds, one of the scattering channels
is an f-channel. Since the two channels must have the same
parity, the second channel is in all likelihood another f-channel, or
a p-channel:  
\begin{itemize}

\item {$\diamond$} $f\otimes f\qquad.$  Candidates: 
Non-Kramers ion, e.g $URu_2Si_2$, $UBe_{13}$.

\item{$\diamond$} $f\otimes p\quad.$   Candidates: $UBe_{13}$,  $UPt_3$ and
Cerium systems, close to quantum critical point. 

\end{itemize}
The first possibility will occur if the
Kondo effect involves a non-Kramer's magnetic ion.
For example, in the case of $URu_2Si_2$, there
is strong circumstantial evidence that the single-ion physics is
dominated by a Kondo effect with a non-Kramers magnetic doublet,
The form factors for the two f-channels in a tetragonal crystal
field are known, and place strong constraints on the 
symmetry of the putative composite order. 
In the second case, the number of available p-channels is small
and the local quantum chemistry will determine the most
likely channel for the co-operative pairing process. 
For example, in hexagonal $UPt_3$, the most likely
second-channel is the $p_z$ orbital, which would explain the presence
of the node in the basal plane. 
In principle, cubic $UBe_{13}$ could belong to either category, 
as this system may also have a non-Kramer's ground-state. If however,
the driving force derives from a p-channel, 
molecular orbital theory dictates that 
the most likely second-channel
is a p-wave state with normal orientated along the cube diagonals,
such as the 111 direction. A gap node normal to this direction
could be detected using
careful transverse ultrasound measurements.\cite{moreno2}

\item One of the paradoxical features of heavy fermion superconductors,
is that their large entropy of condensation suggests a large superconducting
order parameter. Yet Josephson tunneling with a conventional superconductor
has not, to date been achieved. 
Although composite and normal pairs co-exist side-by-side in our
hypothetical superconductor, the predominantly composite character
of the order parameter may help explain why it has proven so
interminably difficult to carry out Josephson tunneling into these
systems. One way to enhance the Josephson current may be to introduce
rare earth, or actinide spins into the tunnel junction.  
Josephson tunneling between a conventional, and composite paired
superconductor requires that the addition of a pair is accompanied by
a spin-flip. Spin fluctuations of the local moments in the junction may help to
catalyse this co-operative process. This is a  possibility currently
under investigation. 

\end{itemize}

We should like to end with a short note about the non-superconducting
aspects of the composite Kondo effect. In our key identity
\bea
{\cal V}\dg_2{\cal V}_1 = - \frac{J_1J_2}{2}
\left[
\matrix{
F &  \Lambda\cr
- \Lambda \dg & F\dg 
}
\right],
\eea
we have the possibility of finite diagonal components $F\ne 0$ due to 
co-operative interference. Unlike composite pairing, such instabilities
will require $J_1$ and $J_2$ to be of comparable size. There are, to our
knowledge two good candidates for this kind of phenomenon:

\begin{itemize}

\item Orbital magnetism in $URu_2Si_2$. As mentioned above,
this material is a naturally occurring two-channel Kondo lattice, but
with strong spin-orbit coupling.  One of the long-standing mysteries
of this compound, is the appearance of an unidentified magnetic
state at $17K$, with a large order parameter which appears to break
time-reversal symmetry, but without producing a large magnetic moment.
\cite{buyers,ramirez}
One possible way to account for this, is to suppose that the two
channels in this compound give rise to a complex order parameter
\bea
F(x) = F_o e^{i \bf Q\cdot \bf x},
\eea
where $\bf Q$ is  commensurate with the lattice.
Just as superconducting composite order coexists with a weak
BCS order parameter, orbital composite
order will co-exist with a weak orbital moment. 
Spin-orbit coupling will then generate a weak magnetic moment. 

\item Ultra-narrow gap Kondo insulators, $CeRhSb$ and $CeNiSn$. These
Kondo insulators appear to develop gap nodes in their tiny hybridization
gap. In a recent paper,\cite{moreno} 
we have pointed out that this kind of behavior
would arise from  the suppression of shape
fluctuations, which gives rise to
three orbital scattering channels in which the Kondo effect can take place.
The resulting interference between the three orbital channels is found
to spontaneously generate a crystal field environment
that gives rise to a Kondo ``insulator'' with gap nodes.

\end{itemize}
These are both areas of active investigation, which lie outside the scope
of this paper. \\

\noindent{\bf Acknowledgments} We should 
like to thank N. d'Abrumenil, A. Georges, R. Ramazashvili and
A. Finkelstein for valuable discussions relating to this work.
Research was supported in part by the National Science Foundation
under Grants NSF DMR 96-14999 , the EPSRC, UK and NATO grant
CRG. 940040. HYK is grateful for support from a Korea Research
Foundation grant during the period when this work was carried out at
Rutgers University. P.C. is grateful for the support of the Oxford
University Physics department, during a sabbatical stay in 1997, when
the early part of this work was carried out.

\vskip 0,4truein
\centerline{\bf APPENDIX A.}
\vskip 0.3 truein
In this section we apply the SU(2) decoupling 
scheme originally developed by 
Affleck et al. for the Heisenberg model.
to
the two-channel Kondo problem.  
This derivation
is closer in spirit to the original work
by Affleck et al, and differs in detail
from the later work by Andrei and
Coleman. The approach of Affleck et al is more
explicitly SU(2) symmetric and can be naturally
extended to include source terms that couple to
the composite order parameter. There are two
distinct differences between the approaches:
\begin{itemize}

\item The  integration measure
over the SU(2) field is "flat"

\item The Gaussian
coefficient of ${\rm Tr}[{\cal V}\dg_{\Gamma}{\cal
V}_{\Gamma}]$ is now $1/J$ rather than $1/2J$, as
it was in the earlier work by Andrei and Coleman.

\end{itemize}

The difference in measure leads to
differences in the fluctuations around
the mean-field theory, and the
mean-field expressions for the Kondo
temperature obtained in the two
methods
actually differ by a factor of two in the
exponential. We have chosen the
 approach of
Affleck et al because it  gives us a much cleaner
an symmetric derivation of the final results.

The objective of
this section is to show how the interaction
between a localized moment $\bf S$ and the 
electron local  spin density
$\sigma_{\Gamma}=\psi\dg\pmb{$\sigma$}\psi$ can be
decoupled  in terms of a fluctuating SU(2) field
\bea
J _{\Gamma}
({\bf S} \cdot \sigma_{\Gamma} -
{\textstyle \frac{1}{2}})
\rightarrow [\tilde  f\dg
{\cal V}_{\Gamma }\tilde \psi_{\Gamma } + H.c ] +
{
 \Tr [{\cal V}_{\Gamma}\dg{\cal
V}_{\Gamma} ]
\over  J_{\Gamma} }\label{k11x}
\eea
where $V_{\Gamma } $ is directly proportional
to an SU(2) matrix $g_{\Gamma }$
\bea
{\cal V}_{\Gamma } = i V^{\Gamma }_{o} g_{\Gamma
}  = \left[\matrix{
V& \Delta\cr
\Delta^*&-V^{ *}}
\right]^{\Gamma}.\eea
For clarity, all site indices $j$ are omitted
from this derivation, but are readily restored
later.

Following earlier work,   we introduce the
following matrix fermions
\bea
{\cal F} = \left[ \matrix{
f_{ \uparrow} & f_{ \downarrow}\cr f^{\dag}_{
\downarrow} & -f^{\dag}_{
\uparrow} }
\right], \quad \Psi _{\Gamma } = \left[ \matrix{
 \psi_{\Gamma \uparrow} &  \psi_{\Gamma
\downarrow}\cr \psi^{\dag}_{\Gamma
\downarrow}&-\psi^{\dag}_{\Gamma \uparrow} }
\right].\eea

By taking the product of these matrix operators
with their Hermitian conjugates, we find that
\bea
\Psi^{\dag}_{\Gamma }
\Psi_{\Gamma}&=&  \underline{1} + 
\underline {\pmb{$\sigma$}}  ^{\rm T}
\cdot\pmb{$\sigma$} _{\Gamma}\cr 
{\cal F}^{\dag}{\cal F} &=& 
\underline{1} + 2
\underline{\pmb{$\sigma$}}^{\rm
T}\cdot{\bf S}\label{thegoods}\eea
where $\pmb{$\sigma$}_{\Gamma} = \psi\dg_{\Gamma }\pmb{$\sigma$}
\psi_{\Gamma }$ is the electron spin density
and $\underline{\pmb{$\sigma$}}^{\rm
T}$ denotes the transpose of the Pauli spin
matrix.
Transformations acting to the right of ${\cal F}$, 
${\cal F} \rarrow {\cal F} h$ 
correspond to physical rotations of the local
moment.  Transformations acting to the left
of ${\cal F}$, ${\cal F}\rarrow g {\cal F}$ correspond to the local
SU(2) transformation, under which the spin
operator is explicitly invariant:
\bea
{\bf S} =\frac{1}{4}
{\rm Tr} [ \pmb{$\sigma$}^T
{\cal F}^{\dag}{\cal F}]\rarrow  \frac{1}{4}
{\rm Tr} [ \pmb{$\sigma$}^T
{\cal F}^{\dag} g\dg g 
{\cal F}] = {\bf S}
\eea
Multiplying the two equations (\ref{thegoods})
together, and taking the trace we obtain
\bea
\pmb{$\sigma$}_{\Gamma }
 \cdot {\bf S} +{\textstyle \frac{1}{2}} = 
\frac{1}{4} {\rm Tr}[{\cal F}^{\dag}
{\cal F}\Psi^{\dag}_{\Gamma }
\Psi_{\Gamma
}
]
\eea
Anti commuting the conduction electron operator
$\Psi_{\Gamma}$
to the left through the trace, we then find that
\bea
J_{\Gamma} \biggl(\pmb{$\sigma$}
_{\Gamma 
} \cdot {\bf S}
-{\textstyle\frac{1}{2}}\biggr) =  -
\frac{J_{\Gamma}}{4} {\rm Tr}[U^{\dag}_{\Gamma
}U_{\Gamma }
].
\eea
where
\bea
U_{\Gamma } 
={\cal F}\Psi^{\dag}_{\Gamma} =
\left[
\matrix{  - a^{\dag}_{\Gamma } &
b_{\Gamma }\cr
b^{\dag}_{\Gamma }&a_{\Gamma}
}
\right]
\eea
is an anti-unitary matrix and
\bea
a_{\Gamma}&=& \sum_{\si}f\dg_{\si}
\psi_{\Gamma\sigma},
\cr
b_{\Gamma}&=& \sum_{\si}\si f_{-\si}
\psi_{\Gamma\sigma},
\eea
Notice that if we expand the above  interaction, we obtain
\bea
H_I = -\frac{J_{\Gamma}}{4}\bigl[
a^{\dag}_{\Gamma}
a_{\Gamma} 
+ a_{\Gamma}a^{\dag}_{\Gamma}+
b^{\dag}_{\Gamma }b_{\Gamma }+
b_{\Gamma }b^{\dag}_{\Gamma }
\bigr]
\eea
showing that it has been decoupled simultaneously in the 
particle-hole, and Cooper channels. 

 We now
apply a  Hubbard-Stratonovich procedure  to this
expression. Formally, we  first convert
each of the fermionic operators in the
interaction to Grassman variables inside
a path integral. On each
time slice we write
\bea
e^{- \Delta \tau H_I} = \int D[{\cal
V}_{\Gamma}, {\cal
V}\dg_{\Gamma}]
e^{-\Delta \tau H_I[{\cal
V}_{\Gamma}, {\cal
V}\dg_{\Gamma}]}
\eea 
where we have transformed
\bea
H_I= 
-\frac{J_{\Gamma}}{4} {\rm Tr}[U^{\dag}_{\Gamma
}U_{\Gamma }
]\longrightarrow H_I[{\cal
V}_{\Gamma}, {\cal
V}\dg_{\Gamma}]\eea
and
\bea
 H_I[{\cal
V}_{\Gamma}, {\cal
V}\dg_{\Gamma}]&=&\cr \frac{1}{2} \biggl\{{\rm Tr}[{\cal
V}^{\dag}_{\Gamma }U_{\Gamma}]&+&
{\rm Tr}[U^{\dag}_{\Gamma}
{\cal V}_{\Gamma }
]\biggr\}+
\frac{1}{J_{\Gamma}}{\rm
Tr}[{\cal V}^{\dag}_{\Gamma }
{\cal V}_{\Gamma
}
].
\eea
A priori, $\cal V$ is a two by two complex
matrix. However, if we divide it up into the
sum of a   unitary and an 
anti-unitary matrix,  we find that only the former
completely decouples. The residual part of $\cal
V$ is completely anti-unitary, and has the form
\bea
{\cal V}_{\Gamma }  = \left[\matrix{
V& \Delta\cr
\Delta^*&-V^{ *}}
\right]_{\Gamma}.
\eea
where there are only two independent complex
parameters. This is a significant simplification.
Notice that ${\cal V}_{\Gamma}$ is directly
proportional to an SU(2) matrix:
\bea
{\cal V}_{\Gamma }= i V^{\Gamma }_{o}
g_{\Gamma } , \qquad  ( V^{\Gamma }_{o}
= \sqrt{|V_{\Gamma }|^2 +|\Delta _{\Gamma }|^2
}).
\eea
The measure of integration for each time-slice is
then simply
\bea
D[{\cal V}_{\Gamma} ,{\cal V}\dg_{\Gamma} ]
=dV_{\Gamma}dV^*_{\Gamma}d\Delta
_{\Gamma}d\Delta^* _{\Gamma}.
\eea
As our final step, we now reduce the
decoupled interaction to a more manageable
two-component notation. Writing
\bea
\tilde f\dg = (f\dg_{\uparrow},
f_{\downarrow}),\cr
\tilde \psi \dg_{\Gamma}=
(\psi\dg_{\Gamma\uparrow},
\psi_{\Gamma \downarrow}),
\eea
then $H_I$ reduces to the form, 
\bea
H_I =
 [\tilde  f\dg
{\cal V}_{\Gamma }\tilde \psi_{\Gamma }+{\rm hc} ]
+ \frac{1}{J_{\Gamma}}{\rm Tr} [{\cal
V}^{\dag}_{\Gamma}{\cal V}_{\Gamma}],
\eea
which is the form quoted in the main text.

\vskip 0.4 truein
\noindent{\bf APPENDIX B.}
\vskip 0.4 truein

The purpose of this section is to establish
the direct relationship 
\bea
{\cal V}\dg_2{\cal V}_1 = - \frac{J_1J_2}{2}
\left[
\matrix{
F\dg &  \Lambda\cr
- \Lambda \dg & F\dg 
}
\right],
\eea
where
\bea
F &= &\psi\dg_2 \pmb{$\sigma$} \psi_1 \cdot {\bf
S},\cr 
{\Lambda} &= &\psi_2 (- i \sigma_y)\pmb{$\sigma$}
\psi_1
\cdot {\bf S}\eea
represent the single composite order in the
particle-hole, and particle-particle channels
respectively. 

In order to establish this identity,
we introduce a source term into the Lagrangian
which couples to the gauge invariant matrix
product ${\cal V}_2{\cal V}_1$, writing
\bea
H_I
&=&\sum_{\Gamma}\biggl\{
 [\tilde  f\dg
{\cal V}_{\Gamma }\tilde \psi_{\Gamma }+{\rm hc} ]
+ \frac{1}{J_{\Gamma}}{\rm Tr} [{\cal
V}^{\dag}_{\Gamma}{\cal V}_{\Gamma}]\biggr\}\cr
&+ &{\rm Tr}
[{\cal V}_2\dg {\cal V}_1 \alpha + h.c ]
\label{start}
\eea
where the source term $\alpha = \alpha_o + i\vec
\alpha
\cdot {\pmb{$\tau$}} $ is a unitary matrix, with
four real coefficients. 

We shall now invert the Hubbard Stratonovich
transformation, with the source terms in place.
We begin by rewriting the Gaussian term
in the interaction to obtain
\bea
H_I
=\sum_{\Gamma}{\textstyle\frac{1}{2}} [{\rm
Tr}[{\cal V}^{\dag}_{\Gamma }U_{\Gamma}]+{\rm hc}
]+ {\rm Tr}
[{\cal V}_{\Gamma }\dg {\cal
V}_{\Gamma'}m_{\Gamma'\Gamma}]
\eea
where
\bea
m_{\Gamma \Gamma'} 
=\left[
\matrix{
{\textstyle \frac{1}{J_1} }& \alpha\cr
\alpha\dg &{\textstyle \frac{1}{J_2}} }\right]
\eea
When we carry out the Gaussian integral
over ${\cal V}_{\Gamma}$, the transformed
Hamiltonian now becomes
\bea
H_I = -\frac{1}{4}{\rm Tr} \bigl[
U\dg_{\Gamma} U_{\Gamma'} (m^{-1})_{\Gamma'\Gamma}
\bigr],
\eea
where
\bea
m^{-1}
={ \frac{J_1J_2}{(1- \vert \alpha\vert ^2 J_1J_2)}}
\left[
\matrix{
{\textstyle \frac{1}{J_2} }& -
\alpha\cr
-\alpha\dg &{\textstyle \frac{1}{J_1} } }\right]
\eea
When we expand this to linear order in $\alpha$
we obtain
\bea
H_I = 
- \sum_{\Gamma} \frac{J_{\Gamma}}{4}
{\rm Tr}[ U\dg_{\Gamma} U_{\Gamma}] 
+ \frac{J_1J_2}{4}{\rm Tr}[U\dg_2 U_1 \alpha +
{\rm hc}]
\label{finish}
\eea
Inserting ${\cal F}\dg {\cal F} =
\underline{1} + 2 \pmb{$\sigma$}^{\rm T}
\cdot {\bf S}$ into this expression,
and making the observation that
\bea
{\rm Tr}\bigl[
\alpha \Psi\dg_2\Psi_1 + {\rm hc}\bigr] = 0
\eea
we can rewrite
\bea
{\rm Tr}[U\dg_2 U_1 \alpha +
{\rm hc}]={2}
{\rm Tr}[\alpha \Psi_2 (\pmb{$\sigma$}^{\rm T}
\cdot {\bf S})\Psi\dg_1  +
{\rm hc}]
\eea
so that the final form of the interaction,
with the source term is
\bea
H_I &=& 
- \sum_{\Gamma} \frac{J_{\Gamma}}{4}
{\rm Tr}[ U\dg_{\Gamma} U_{\Gamma}] \cr
&+& \frac{J_1J_2}{2}{\rm Tr}[\alpha \Psi_2
(\pmb{$\sigma$}^{\rm T}
\cdot {\bf S})\Psi\dg_1 
 +
{\rm hc}]
\label{finito}
\eea
comparing  coefficients of $\alpha$
in (\ref{start}) with (\ref{finito}), we obtain
 the following identity 
\bea
{\cal V}\dag_2{\cal V}_1 =
\frac{J_1 J_2}{2}[\Psi\dg _2(\pmb{$\sigma$}^{\rm
T}
\cdot {\bf S}) \Psi\dg_1]\label{mainr}
\eea
This expresses, in a compact form, the
relationship between the inter-channel
interference and the composite order.
To complete the job,
we now expand the right-hand side.  
We first write
\bea
[\Psi _2(\pmb{$\sigma$}^{\rm
T}
\cdot {\bf S}) \Psi\dg_1] 
=
-[\Psi _2
(\sigma_y
\pmb{$\sigma$}\sigma_y
\cdot {\bf S}) \Psi\dg_1] 
\eea
where we have replaced $\pmb{$\sigma$}^{\rm T}
= - (\sigma_y
\pmb{$\sigma$}\sigma_y)$.
To make the expansion, it
it is convenient to write
\bea
\Psi_{2}= \left(\matrix{ \psi_{2}^{\rm T}
\cr
\psi\dg _{2} (i \sigma_y)}\right), \quad
\Psi\dg _{1}= \left( \psi_{1}^{*}, 
-i \sigma_y\psi _{1} \right)
\eea
where $\psi_{2}^{\rm T}$ is the
row-spinor formed by taking the transpose of the
column spinor
$\psi_2$, 
and $\psi_{1}^{*}= (\psi\dg_1)^{\rm T}$  is the
column spinor formed by taking the transpose
of $\psi\dg_1$. 
Multiplying out the 
matrices, we obtain
\bea
-\frac{2}{J_1J_2}{\cal V}\dag_2{\cal V}_1 
&=& 
\left[\matrix{
\psi^T_2 \sigma_y\pmb{$\sigma$}\sigma_y\psi^*_1 & 
\psi^T_2 (-i \sigma_y \pmb{$\sigma$})\psi_1\cr
\psi\dg_2(\pmb{$\sigma$}i \sigma_y )\psi^*_1 &
\psi_2\dg  \pmb{$\sigma$}\psi_1
&  }
\right]\cdot {\bf S}\cr
&= &
\left[\matrix{
\psi_1\dg  \pmb{$\sigma$}\psi_2 & 
\psi_1 (i \sigma_y \pmb{$\sigma$})\psi_2\cr
\psi\dg_2(\pmb{$\sigma$}i \sigma_y )\psi^*_1 &
\psi_2\dg  \pmb{$\sigma$}\psi_1
&  }
\right]\cdot {\bf S}\cr
&=&\left[
\matrix{
F &  \Lambda\cr
- \Lambda \dg & F\dg 
}
\right].\label{bigjob}
\eea
Substituting (\ref{bigjob}) into (\ref{mainr}))
we obtain the quoted result.

\vskip 0.4 truein
\noindent{\bf APPENDIX D.}
\vskip 0.4 truein

The purpose of this section is to evaluate the susceptibilities
associated with the expansion of the mean-field Free energy about
the pure composite paired state 
\bea
F= F_o - \frac{1}{2}\bigl[ \chi_{\mu \mu} (\delta \mu)^2 +  
2\chi_{\mu \lambda}\delta \mu \delta \lambda + \chi_{\lambda \lambda}
(\delta \lambda )^2 \bigr].
\eea
By integrating over the Gaussian
fluctuations in $\lambda$ to impose the constraint on the f-charge, 
we can use these susceptibilities
to compute the physical charge susceptibility 
\bea
\chi_C = \chi_{\mu \mu} - \frac{ (\chi_{\mu \lambda})^2 }{  \chi_{\lambda \lambda}   }.
\eea

To compute the susceptibilities we expand the Hamiltonian about the
half-filled state, 
\def\zap{\textstyle\left[\matrix{\tau_3 & 0 \cr
0 &0}\right]
}
\def\zad{\textstyle\left[\matrix{0 & 0 \cr
0 &\tau_3}\right]
}
\bea
{\cal H} = {\cal H}_o - \delta \mu \zap + \lambda \zad
\eea
The electronic part of the Free energy is given by
\bea
F_e = -T \sum_{\kappa}{\rm Tr} {\rm ln} [ i \omega_n - {\cal H}({\bf k})]
\eea
Expanding this to second-order then gives
\bea
\chi_{\mu \mu} &=& - T \sum_{\kappa} {\rm Tr} \biggl(
{\cal G}_{\kappa}\zap{\cal G}_{\kappa}\zap
\biggr)\cr
\chi_{\mu \lambda} &=&  T \sum_{\kappa} {\rm Tr} \biggl({\cal G}_{\kappa}\zap{\cal G}_{\kappa}\zad
\biggr)\cr
\chi_{\lambda \lambda} &=& - T \sum_{\kappa} {\rm Tr} \biggl(
{\cal G}_{\kappa}\zad{\cal G}_{\kappa}\zad
\biggr)
\eea
At half-filling, the electron propagator can be written
\bea
{\cal G}(\kappa) = 
\left[\matrix{ \omega \tilde G &- i v_{\bf k}\tilde G g \dg \cr
  i v_{\bf k}g\tilde G & g(\omega -
\epsilon_{\bf k} \tau_3 )\tilde G g\dg
 }
\right]_{\kappa}
\eea
where $v_{\bf k} =\sqrt{v_{1 {\bf k}}^2 +v_{2 {\bf k}}^2 }$,
\bea
\tilde G(\kappa) = \frac{1}{\omega ( \omega- \epsilon_{\bf k}\tau_3) -
v_{\bf k}^2  }
\eea
and
\bea
g =  \frac{1}{v_{\bf k}}
\left[ v_{1 {\bf k}} -i v_{2 {\bf k}}\tau_2 \right]
\eea
The Green function ${\cal G}$  has poles at $\pm E_{\bf k \eta}$
where
\bea
E_{\bf k\eta } = \frac{\epsilon_{\bf k}}{2}
+ \eta \sqrt{ 
\bigl(
\frac{\epsilon_{\bf k}}{2}\bigr)^2+ v_{\bf k}^2}, \qquad (\eta = \pm)
\eea
Expanding the electron propagator about its poles, we write
\bea
\tilde G(\kappa) &=& \sum_{\eta = \pm} G_{\eta}(\kappa)
\frac{ \eta }{\sqrt{
\epsilon_{\bf k}^2 + 4 v_{\bf k}^2}}\tau_3\cr
G_{\eta}(\kappa)&=&
\frac{1}{\omega - E_{{\bf k}\eta}\tau_3} .
\eea
Inserting this into the full propagator, we can write it in the form
\bea
{\cal G}(\kappa) &=& \sum_{\eta}
\left[\matrix{ c_{\bf k \eta}^2 G_{\eta}&  -i c_{\bf k \eta}s_{\bf k \eta} G_{\eta} \tau_3 g\dg\cr
i c_{\bf k \eta}s_{\bf k \eta}g\tau_3 G_{\eta}  & s_{\bf k \eta}^2 g G_{\eta} g\dg 
 }
\right]_{\kappa}\cr
&=& \sum_{\eta} \zeta_{ {\bf k}\eta} \otimes G_{\eta}(\kappa)
\zeta\dg_{ {\bf k}\eta}
\eea
where
\bea
\zeta_{ {\bf k}\eta}= \left(
\matrix{
c_{{\bf k}\eta}\cr
i s_{{\bf k}\eta} g_{\bf k} \tau_3}
\right)\eea
is the eigenvector corresponding to the quasiparticle with energy
$E_{\bf k \eta}$. The quantities
\bea
c^2_{\bf k\eta}&=& 
\frac{E_{\bf k\eta}}
{
E_{\bf k \eta} -
E_{\bf k -\eta}}= 
\frac{1}{2}\left[ 1 + \eta\frac{\epsilon_{\bf k}}
{\sqrt{
\epsilon_{\bf k}^2 + 4 v_{\bf k}^2}}
\right]\cr
s^2_{\bf k\eta}&=&
\frac{E_{\bf k\eta}- \epsilon_{\bf k}}
{
E_{\bf k \eta} -
E_{\bf k -\eta}}= 
 \frac{1}{2}\left[ 1 - \eta\frac{\epsilon_{\bf k}}
{\sqrt{
\epsilon_{\bf k}^2 + 4 v_{\bf k}^2}}
\right]\cr
c_{\bf k\eta}s_{\bf k\eta}&=& 
\frac{v_{\bf k}}
{
E_{\bf k \eta} -
E_{\bf k -\eta}}= 
\eta \frac{v_{\bf k}}{
\sqrt{
\epsilon_{\bf k}^2 + 4 v_{\bf k}^2}}
\eea
describe the admixture between the conduction, and f-electrons.
The matrix elements of the charge operators appearing inside
the susceptibilities are
\bea
\zeta_{\bf k \eta}\dg 
\zap
\zeta_{\bf k \eta'}&=& c_{\bf k \eta}c_{\bf k \eta'} \tau_3\cr
\zeta_{\bf k \eta}\dg 
\zad
\zeta_{\bf k \eta}&=&s_{\bf k \eta}s_{\bf k \eta'}
\tau_3g_{\bf k}^2 \cr
&= &s_{\bf k \eta}s_{\bf k
\eta'} ( \cos \phi_{\bf k} \tau_3 + \sin
\phi_{\bf k}
\tau_2)
\eea
where $C_{\bf k} = v_{\bf k - }^2 /
v_{\bf k}^2$ and $S_{\bf k} = 2 v_{1 \bf
k}v_{2\bf k}/v_{\bf k}^2
$. The expressions for the
susceptibilities can now be written
\bea 
\chi_{\mu \mu}&=& -T\sum_{\kappa, \eta ,  \eta'}
{\rm Tr}\biggl[
G_{\eta} \tau_3 G_{\eta'} \tau_3
\biggr]\cr
\chi_{\mu \lambda}&=& T\sum_{\kappa, \eta , 
\eta'} {\rm Tr}\biggl[
G_{\eta} \tau_3 G_{\eta'}
\tau_3
\biggr]C_{\bf k}
\label{itt}\cr
\chi_{\lambda\lambda}&=& -T\sum_{\kappa, \eta
, 
\eta'} {\rm Tr}\biggl[
G_{\eta} \tau_3 G_{\eta'}
\tau_3
C_{\bf k}^2
 +
G_{\eta}\tau_2 G_{\eta'} 
\tau_2S_{\bf k}^2
\biggr]\nonumber
\eea
where we 
denote
$G_{\eta}\equiv G_{\eta}(\kappa)$ and vanishing
cross-terms between $\tau_3$ and $\tau_2$ have
been dropped. 
We now evaluate the Matsubara sums in these
expressions, and take the zero-temperature
limit. Key results that we use are
\bea
-T \sum_{\i \omega_n,  \eta'}
{\rm Tr}\biggl[
G_{\eta} \tau_3 G_{\eta'} \tau_3
\biggr]  c^2_{\eta}c^2_{\eta'}&= & 2\sum _{
\eta'}
\frac{ f_{\bf k \eta'}-f_{\bf k \eta}
}{E_{\bf k\eta}
- E_{\bf k \eta'}}c^2_{\eta}c^2_{\eta'}\cr
&\rarrow&\frac{ 2 c^2 _{\eta} s^2
_{\eta}}{\sqrt{\epsilon_{\bf k}^2 + 4 v_{\bf k}^2}
},\nonumber\eea
where $f_{\bf k \eta} \equiv 1/(e^{\beta E_{\bf k \eta}}+1)$ denotes the Fermi function. Similarly,
\bea
T \sum_{\i \omega_n,  \eta'}
{\rm Tr}\biggl[
G_{\eta} \tau_3 G_{\eta'} \tau_3
\biggr]  ( c s)_{\eta} ( c
s)_{\eta'}&\rarrow&\frac{ 2 c^2 _{\eta} s^2
_{\eta}}{\sqrt{\epsilon_{\bf k}^2 + 4 v_{\bf
k}^2} },\cr
T \sum_{\i \omega_n,  \eta'}
{\rm Tr}\biggl[
G_{\eta} \tau_3 G_{\eta'} \tau_3
\biggr]   s^2_{\eta}  s^2
_{\eta'}&\rarrow&\frac{ 2 c^2 _{\eta} s^2
_{\eta}}{\sqrt{\epsilon_{\bf k}^2 + 4 v_{\bf
k}^2} }.\nonumber
\eea
Lastly, there is one anomalous term, 
\bea
-T \sum_{\i \omega_n,  \eta,\eta'}
{\rm Tr}\biggl[
G_{\eta} \tau_2 G_{\eta'} \tau_2
\biggr]  s^2_{\eta}s^2_{\eta'}& \rarrow&\sum_{\eta}\frac{  s^4 _{\eta}
}{|E_{\bf k \eta}| }\cr &=&
\frac{ v_{\bf k}^2 +  \epsilon_{\bf k}^2 }{v_{\bf k}^2\sqrt{\epsilon_{\bf k}^2 + 4 v_{\bf k}^2} }.\nonumber
\eea
Putting these results together, we obtain
\bea
\chi_{\mu \mu} &=&
\sum_{\bf k}
\frac{ 4 v_{\bf k}^2 }{(\epsilon_{\bf k}^2 + 4
v_{\bf k}^2)^{3\over 2}}\cr
\chi_{\mu \lambda}  &=&
\sum_{\bf k}
\frac{ 4 v_{\bf k -}^2 }{(\epsilon_{\bf k}^2 + 4
v_{\bf k}^2)^{3\over 2}}\cr
\chi_{\lambda \lambda} &=& \sum_{\bf k}
\frac{ 4 v_{\bf k}^2 }{(\epsilon_{\bf k}^2 + 4
v_{\bf k}^2)^{3\over 2}}\left( \frac{v_{{\bf k}-}^2}{v_{\bf k}^2}\right)^2
\cr
&+& \sum_{\bf k} \frac{ v_{\bf k}^2 + \epsilon_{\bf k}^2}
{v_{\bf k}^2\sqrt{\epsilon_{\bf k}^2 + 4 v_{\bf k}^2}}
\left({2 v_{1 {\bf k}}v_{2{\bf k}}\over 
 v_{\bf k }^2}
\right)^2 
\eea
We can also rewrite $\chi_{\lambda\lambda}$ in the following form
\bea
\chi_{\lambda \lambda} &=& \chi_{\mu \mu} + \chi_b\cr
\chi_b &=& \sum _{\bf k} \frac{\epsilon_{\bf k}^2 }{[(\epsilon_{\bf k})^2
+ 4v_{{\bf k}}^2]^{3/2}}\left(\frac{ 2 v_{1{\bf k}}v_{2{\bf k}}}
{v_{{\bf k}}^2}
\right)^2({5} + \frac{\epsilon_{\bf k}^2 }{v_{\bf k}^2}).
\eea
Since $\chi_{\lambda \lambda} \ge \chi_{\mu \mu}$, and
$|\chi_{\mu \lambda}| \le \chi_{\mu \mu}$, the 
the final result for the charge susceptibility is then guaranteed
to be positive when $v_1v_2 \ne 0$:
\bea
\chi_C = \chi_{\mu\mu} - \frac{(\chi_{\mu \lambda})^2}{\chi_{\mu \mu}+ \chi_b}
>0, \qquad (v_1v_2 \ne 0).
\eea

\end{document}